\documentclass[a4paper,11pt]{report}
\usepackage[T1]{fontenc}
\usepackage[utf8]{inputenc}
\usepackage{amsfonts,amsmath,amssymb}
\usepackage[left=25mm,right=20mm,bottom=25mm,top=25mm,headheight=10mm]{geometry}
\usepackage{lastpage}
\usepackage{bm}
\usepackage{xcolor}
\usepackage[nottoc]{tocbibind}
\usepackage{titletoc}
\usepackage{fancyhdr}
\usepackage{booktabs}
\usepackage{graphicx}
\usepackage{hyperref}
\usepackage{enumitem}
\usepackage[font=small]{caption}
\usepackage{fontawesome}
\usepackage{authblk}
\usepackage{titlesec}
\usepackage{tikz,tcolorbox}
\usepackage[toc,page]{appendix}
\usepackage{apptools}
\usepackage[
   natbib=true,
    style=numeric,
    bibstyle=ieee,
    citestyle=numeric-comp,
    sorting=none]{biblatex}
\usetikzlibrary{calc, intersections, arrows}
\dottedcontents{section}[2.5em]{}{1.5em}{0.5pc}
\dottedcontents{subsection}[5.5em]{}{2em}{0.5pc}
\numberwithin{equation}{section}

\hypersetup{linktoc = all}

\titleformat{\chapter}[display]
  {\normalfont\huge\bfseries\filcenter}{\chaptertitlename\ \thechapter}{20pt}{\Huge}
\titlespacing*{\chapter}
  {0pt}{30pt}{20pt}

\titleformat*{\section}{\Large\bfseries}
\titleformat*{\subsection}{\large\bfseries}

\counterwithout{section}{chapter}

\makeatletter
\newcommand{\customlabel}[2]{%
   \protected@write \@auxout {}{\string \newlabel {#1}{{#2}{\thepage}{#2}{#1}{}} }%
   \hypertarget{#1}{#2}
}
\makeatother

\definecolor{lime}{HTML}{A6CE39}
\DeclareRobustCommand{\orcidicon}{
   \begin{tikzpicture}
   \draw[lime, fill=lime] (0,0) 
   circle [radius=0.16] 
   node[white] {{\fontfamily{qag}\selectfont \tiny ID}};
   \draw[white, fill=white] (-0.0625,0.095) 
   circle [radius=0.007];
   \end{tikzpicture}
   \hspace{-2mm}
}

\foreach \x in {A, ..., Z}{\expandafter\xdef\csname orcid\x\endcsname{\noexpand\href{https://orcid.org/\csname orcidauthor\x\endcsname}
         {\noexpand\orcidicon}}
}

\definecolor{bluish}{HTML}{228CEE}
\definecolor{greenish}{HTML}{419127}
\definecolor{mulberry}{HTML}{770737}
\definecolor{orangish}{HTML}{FFB347}
\hypersetup{%
  colorlinks=true,
  linkcolor=bluish,
  anchorcolor=green,
  citecolor=orangish,
  filecolor=cyan,
  menucolor=magenta,
  runcolor=yellow,
  urlcolor=greenish
}

\title{\bf\textsc{Lectures on Gravitational Wave Signatures \linebreak of Primordial Black Holes}}
\author[$\ddagger$\orcidA{}]{Guillem Dom{\`e}nech\footnote{\faEnvelope: \href{mailto:{guillem.domenech}@{itp.uni-hannover.de}}{{guillem.domenech}@{itp.uni-hannover.de}}\quad;\quad \faHome: \href{https://domenechcosmo.netlify.app}{domenechcosmo.netlify.app}}}
\affil[$\ddagger$]{Institute for Theoretical Physics, Leibniz University Hannover, \linebreak Appelstraße 2, 30167 Hannover, Germany.}
\date{\today}
\begin{document}
\vspace*{-8cm}
{\let\newpage\relax\maketitle}
\thispagestyle{empty}
\vspace*{-6cm}
\begin{center}
\noindent\textbf{Lectures notes prepared for the ICCUB School 2023 on Primordial Black holes \linebreak in the University of Barcelona}
\end{center}
\vspace*{5mm}

\section*{Abstract}
We provide a pedagogical approach to gravitational waves in cosmology with focus on gravitational wave signals related to primordial black holes. These lectures notes contain more details than one is able to present in the two two-hour lectures they are meant to and, as such, they should be thought as a complementary material. The main aim of these lectures is that, by the end, one obtains a certain degree of intuition on gravitational waves in cosmology and understands the basic features of scalar induced gravitational waves. We also highlight must-check properties of induced gravitational waves as well as current issues regarding secondary gravitational waves in cosmology. Throughout the lecture we provide exercises, supplementary information and activities with public codes to be ready to derive your own results.

\tableofcontents

\chapter*{Lecture 1}
\addcontentsline{toc}{chapter}{Lecture 1}
\markboth{Lecture 1}{Lecture 1}

\section{Introduction \label{sec:intro}}

It is an exciting time to study early universe cosmology with Gravitational Waves (GWs). Present GW interferometers (LIGO/VIRGO/KAGRA) are detecting more and more GW events as time goes by, some of which could in principle be Primordial Black Holes (PBHs). Pulsar Timing Arrays (PTA) are showing tentative hints of a GW background, which could be of cosmological origin. 

There are also future GW detectors planned (such as LISA, Einstein Telescope and Cosmic Explorer) which will greatly improve the sensitivity. And, as sensitivity is improved, no one knows what we will discover. If you do astrophysics, the general expectation is that astrophysical signals will be louder than cosmological ones. But, the truth is we do not know yet. And, even if cosmological signals are subdominant, there is hope that by using information from astrophysics one may dig up the cosmological component.

PBHs and their associated GWs are one of the promising signals that, if detected, will provide groundbreaking insight on the physics of the very early universe. After a brief introduction to the history of our universe (so that we know which periods we consider), the lectures are divided into:
\begin{itemize}
\item[\S~\ref{sec:2}] The basics of GWs in cosmology,
\item[\S~\ref{sec:3}] Understanding GWs signals associated to PBHs,
\item[\S~\ref{sec:4}] Grasp the physics behind scalar induced GWs.
\end{itemize}

\subsection{Brief recap of the history of the Universe}

I am sure that if you have ever attended a cosmology course or heard/watched a cosmology related talk, you will have a good idea of what we think happened in the early universe. Nevertheless, I would like to briefly put my two cents in the story.

Let me start from the very beginning (maybe there was something before, we just don't know). From Cosmic Microwave Backgound (CMB) observations we have pretty good evidence that there was a period of cosmic inflation in the very early universe. Cosmic inflation is a phase of exponential expansion of the universe. Most importantly, quantum fluctuations during cosmic inflation become the seeds of all the structure, like galaxies, we observe today in our Universe. Long story short, during inflation, short wavelength quantum \textit{vacuum} fluctuations  are stretched to the point they become long wavelength; short and long as compared to the Hubble radius during inflation, which is mostly constant. So, cosmic inflation ends with many of what we call super-horizon fluctuations (or super-Hubble-horizon or super-Hubble-radius depending on how rigorous you want to be). Super-Hubble fluctuations are out of causal contact and, therefore, they don’t evolve. That is, until they are in causal contact again, once they become once again sub-Hubble fluctuations.\\

\begin{figure}
\centering
\includegraphics[width=0.6\columnwidth]{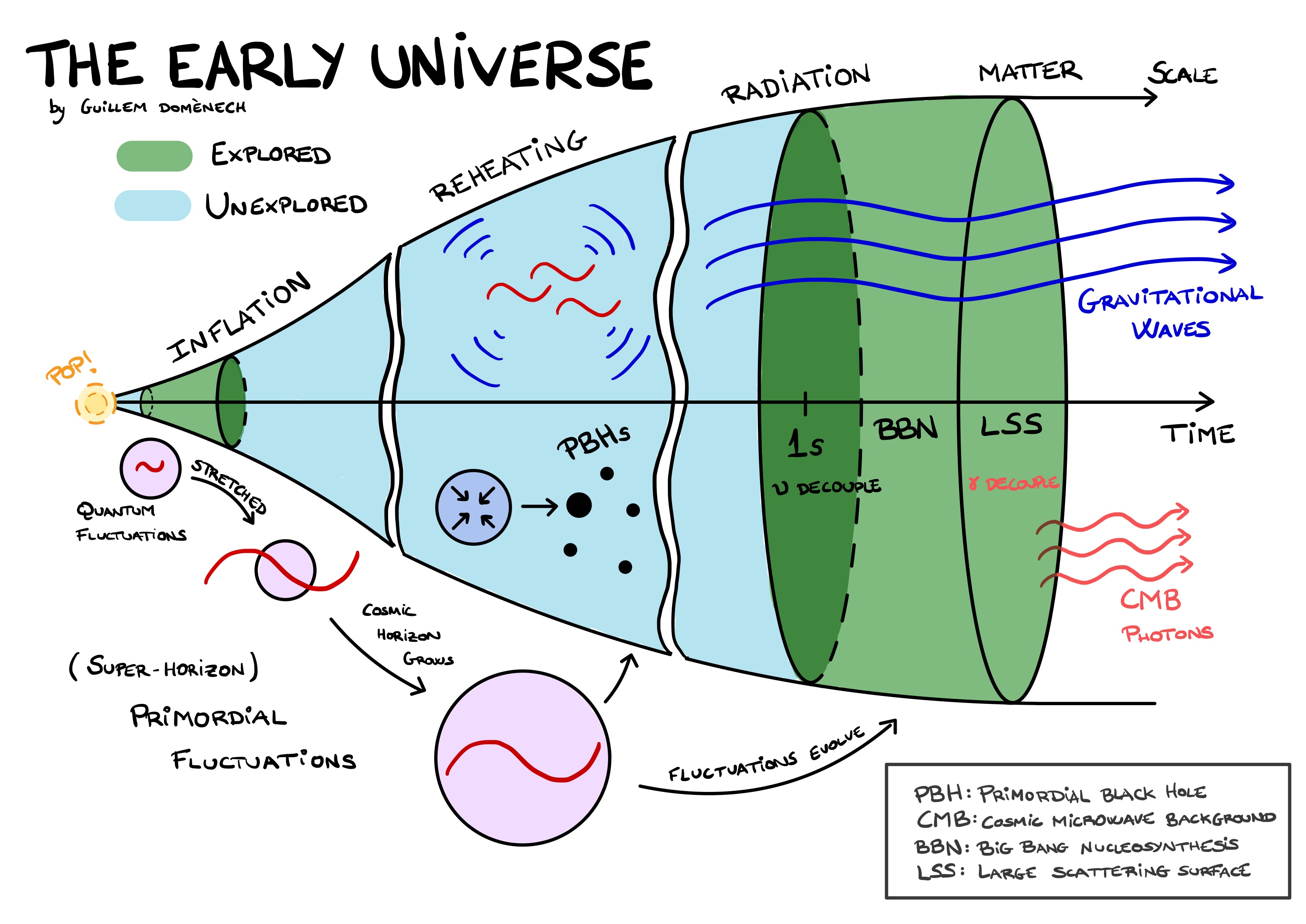}
\caption{Illustration of the history of the Universe. \label{fig:history}}
\end{figure}

What I would like to emphasize is that our current information of the very early universe is mainly based on the CMB, i.e. photons that propagated freely after decoupling from electrons (which occurred in the dark matter dominated era). Unfortunately, before photon decoupling the universe was “opaque” to photons. Nevertheless, with the CMB we have a good idea of the universe since roughly the time of neutrino decoupling ($T\sim \rm MeV$, $t\sim 1\,\rm s$). Also, with CMB temperature (and polarization) fluctuations we probe the first observable e-folds of inflation. But, as illustrated in Fig.~\ref{fig:history}, there is a big chunk of inflation and until $T\sim \rm MeV$ that we have not directly explored yet. Here is where cosmology with GWs becomes very exciting. GWs generated during these unexplored periods (during inflation or after) can in principle reach us. If so, we could find evidence of wild events in the early universe \cite{Caprini:2018mtu}: phase transitions, cosmic strings, primordial black holes, etcetera.\\

\begin{tcolorbox}[title=\bf{Supplementary information},coltitle=black,colback=white,colframe=mulberry!50]
Scalar field fluctuations in an accelerated expanding Universe develop a non-trivial root mean squared when their wavelength become super-Hubble. So if we call the field $\phi$ and its fluctuations $\delta\phi$, we have that the mean $\langle\delta\phi\rangle=0$ but $\sqrt{\langle\delta\phi^2\rangle}\sim H$ (the Hubble radius). However, CMB initial temperature fluctuations are related to the so-called curvature perturbation ${\cal R}$, which is related to $\delta\phi$ by
\begin{align}
{\cal R}\sim \frac{H}{\dot \phi}\delta\phi\sim \frac{H}{\sqrt{\epsilon}}\,,
\end{align}
where $\epsilon$ is the so-called first slow-roll parameter.
In the equation above everything should be thought of as root mean squared. The curvature perturbation is what conveniently sets the initial conditions for primordial fluctuations, as it is constant on super-Hubble scales. From the CMB we have that $\sqrt{\langle {\cal R}^2\rangle}\sim 10^{-4}$. But, to form a significant number of PBHs we need $\sqrt{\langle {\cal R}^2\rangle}\sim 10^{-1}$ and for detectable secondary GWs we need $\sqrt{\langle {\cal R}^2\rangle}\sim 10^{-2}$. So if we see PBHs and/or GWs there was definitely a departure from standard inflation (with small $\epsilon$).
\end{tcolorbox}

\subsection{List of recommended references}

There are many great and detailed lectures online. In these couple of lectures we will not dwell into all the details of the calculations. Priority will be placed on conceptual aspects and currently open topics and issues. Those interested in following more details in the calculations can find exhaustive sources in the list of references below.\\

\noindent On more \textbf{general} topics:
\begin{itemize}

\item[\faBook] Michele Maggiore’s book: “Gravitational Waves: Volume 1: Theory and Experiments”. For gravitational waves basics.

\item[\faBook] Viatcheslav Mukhanov’s book: “Physical Foundations of Cosmology”.  For cosmology basics.

\item[\faNewspaperO] \href{https://arxiv.org/pdf/gr-qc/0501041.pdf}{“The basics of gravitational wave theory”} by Flanagan and Hughes. More on gravitational wave astronomy.

\item[\faNewspaperO] \href{https://arxiv.org/pdf/1801.04268.pdf}{“Cosmological Backgrounds of Gravitational Waves”} by Caprini and Figueroa. More on gravitational wave cosmology.

\item[\faPencilSquareO] \href{http://cosmology.amsterdam/education/cosmology/}{“Lecture notes on Cosmology”} by Daniel Baumann. These are very good for general early universe cosmology.

\item[\faPencilSquareO] Chiara Caprini Youtube’s lectures on GWs \href{https://youtu.be/Pt1rZJCKIsI}{\faYoutubePlay}.

\item[\faPencilSquareO] Valerie Domcke Youtube’s lectures on GWs \href{https://youtu.be/nnCWhSoxGKk}{\faYoutubePlay}. Specialized to axion inflation towards the end.
\end{itemize}

\noindent On more \textbf{specialized} topics:
\begin{itemize}
\item[\faNewspaperO] \href{https://arxiv.org/pdf/1801.05235.pdf}{
“Primordial Black Holes
- Perspectives in Gravitational Wave Astronomy -”} by Sasaki, Suyama, Tanaka and Yokoyama. On PBHs in general (including formation and mergers).
\item[\faNewspaperO] \href{https://arxiv.org/pdf/2002.12778.pdf}{“Constraints on Primordial Black Holes”} by Carr, Kohri, Sendouda and Yokoyama.
\item[\faNewspaperO] \href{https://arxiv.org/pdf/2109.01398.pdf}{“Scalar Induced Gravitational Waves Review”} by myself. Focus on scalar induced GWs.
\item[\faCopy] \href{https://arxiv.org/pdf/2004.04740.pdf}{“Dark matter and dark radiation from evaporating primordial black holes”} by Isabella Masina. Not a review article but you can find all relevant formulas and literature on black hole evaporation.
\end{itemize}

\hrule
\vspace*{2mm}
\noindent \textbf{Legend}:
\begin{center}
\faBook: Book \hspace{5mm} ;\hspace{5mm} \faPencilSquareO: Lectures\hspace{5mm} ;\hspace{5mm}  \faNewspaperO: Review article\hspace{5mm} ;\hspace{5mm}  \faCopy: Research article.
\end{center}
\hrule

\subsection{Warm up your numbers}

In gravity and cosmology, we always encounter the combination ${c^4}/{8\pi G}$. This is what appears in the famous Einstein Equations, $G_{\mu\nu}=\tfrac{c^4}{8\pi G}T_{\mu\nu}$. But keeping track of $c$’s, $\pi$’s and $G$’s is tedious and most likely lead to typos/mistakes. It is much more convenient, to simplify calculations, to work in units where $\hbar=c=1$. However, after doing calculations, it is important to understand the physical meaning of the numbers we obtain. So we should be able to recover units when necessary. To do so, we can compute what are the conversion rules between units for $\hbar=c=1$.

We start by computing the value of the reduced Planck mass, which is defined by
\begin{align}
M_{\rm pl}=\sqrt{\frac{\hbar c}{8\pi G}}\approx 4.34\times 10^{-6}\,{\rm g}\,.
\end{align}
You can check, e.g., \href{https://en.wikipedia.org/wiki/Planck_units}{Wikipedia} for the values of $\hbar$, $c$ and $G$. We can also compute the reduced Planck energy, namely
\begin{align}
E_{\rm pl}=M_{\rm pl}c^2\approx 2.43\times 10^{18}\,{\rm GeV}.
\end{align}
We can also compute the reduced Planck length, temperature, time, etcetera. However, since they are all related by various products of $M_{\rm pl}$, $\hbar$ and $c$, they have to all be proportional to each other when $\hbar=c=1$. We can use that to build a conversion table, like Tab.~\ref{tab:conversion}. For even more convenience, we will also eventually set $M_{\rm pl}^2=1$. But, we can always recover the units when needed. Some other numbers relevant for cosmology are given in Tab.~\ref{tab:quantities}.

\begin{table}[!htbp]
\centering
\begin{tabular}{||c c c c c||} 
 \hline
 Energy & Length$^{-1}$ & Mass & Time$^{-1}$ & Temperature \\ [0.5ex] 
 \hline\hline
 $1\,{\rm eV}$ & $51000\,{\rm cm}^{-1}$ & $1.8\times 10^{-33}\,{\rm g}$ & $1.5\times 10^{15}\,{\rm s}^{-1}$ &$12000\,{\rm K}$  \\ 
 \hline
$M_{\rm pl}\approx2.43\times 10^{18}\,{\rm GeV}$ & $1.2\times 10^{32}\,{\rm cm}^{-1}$ & $4.3\times 10^{-6}\,{\rm g}$ & $3.7\times 10^{42}\,{\rm s}^{-1}$ &$2.8\times 10^{31}\,{\rm K}$ \\[1ex] 
 \hline
\end{tabular}
\caption{Conversion table for natural units. If you do not like this table, \href{https://www.saha.ac.in/theory/palashbaran.pal/conv.html}{this one} by P. B. Pal is better.\label{tab:conversion}}
\end{table}

\begin{table}[!htbp]
\centering
\begin{tabular}{||c c c||} 
 \hline
 Quantity & Notation & Value \\ [0.5ex] 
 \hline\hline
 Comoving wavenumber at equality &  $k_{\rm eq}$  & $0.01\,{\rm Mpc}^{-1}$   \\ 
 \hline
Redshift at equality &  $z_{\rm eq}$ & $3411$ \\
 \hline
Temperature of radiation at equality &$T_{\rm eq}$  & $0.8\,{\rm eV}$ \\ 
\hline
Energy density degrees at equality & $g_*(T_{\rm eq})$  & $3.38$ \\ 
\hline
Entropy density degrees at equality & $g_{*s}(T_{\rm eq})$  & $3.94$ \\ 
\hline
Hubble parameter today & $H_{0}$   & $67\, \rm km/s/Mpc$ \\ 
\hline
Dimensionless Hubble parameter & $h$ & $H_{0}/(100 \, \rm km/s/Mpc)$\\ 
\hline
Temperature of photons today & $T_{0}$ & $2.73 \, \rm K$ \\ 
\hline
Fraction of radiation today&$\Omega_{\rm rad,0}h^2$& $4.18\times 10^{-5}$ \\
 \hline
 Fraction of dark matter today&$\Omega_{\rm cdm,0}h^2$& $0.12$ \\
 \hline
Energy density degrees today & $g_*(T_{0})$ & $3.38$ \\ 
\hline
Entropy density degrees today & $g_{*s}(T_{0})$  & $3.94$ \\ 
\hline
Solar Mass & $M_\odot$ & $2\times 10^{33}\,\rm g$ \\ 
\hline
Megaparsec & $\rm Mpc$ & $3.1\times 10^{24}\,{\rm cm}$ \\ [1ex] 
\hline
\end{tabular}
\caption{Some relevant quantities for cosmology. For more check Table 2 of the Planck 2018 paper on cosmological parameters \cite{Aghanim:2018eyx}. Here $1+z=a_0/a$ where $a$ is the scale factor of the Universe. \label{tab:quantities}}
\end{table}

\clearpage

\section{Gravitational Waves in Cosmology \label{sec:2}}

Let us briefly go through the concepts and formalism that we will use throughout the lectures. Part of this section might be a review of other lectures in the school or well-known to those who took a cosmology course. But, to make things a bit more interesting I will use a different approach to derive the main equations and, hopefully, you can learn some tricks to speed up calculations next time.

To properly understand GWs in a cosmological set up we need to:
\begin{itemize}
\item[\S~\ref{sub:universe}]  Be familiar with the equations describing the Universe,
\item[\S~\ref{sub:GWs?}]  Connect cosmological perturbations to GWs,
\item[\S~\ref{sec:GWsincosmo}]  Understand how GWs behave in an expanding universe,
\item[\S~\ref{sub:GWsenergy}]  Connect predictions of the theory with experiments.
\end{itemize}

\subsection{How do we describe the Universe? \label{sub:universe}}
The universe on scales larger than roughly $100\,{\rm Mpc}$ looks on average homogeneous and isotropic (for comparison a galaxy has a size of $\sim 10\,{\rm kpc}$. Galaxies are separated by $\sim 1\,{\rm Mpc}$). The effect of gravity is then described by General Relativity (although there is a lot of work going on about modifications of GR). The geometry of a homogeneous and isotropic universe is captured by the so-called the Friedmann–Lemaître–Robertson–Walker (FLRW) metric. For us, it is convenient to write it as
\begin{align}\label{eq:FLRW}
\tilde g_{\mu\nu}=a^2(\tau)g_{\mu\nu}=a^2(\tau)\left(\eta_{\mu\nu}+h_{\mu\nu}\right)\,,
\end{align}
where $\eta_{\mu\nu}$ is the Minkowski (flat) metric and I took the liberty to consider some perturbations $h_{\mu\nu}$ on top for generality. Essentially, FLRW is a Minkowski metric that scales with the scale factor $a(\tau)$ over time. $\tau$ is called the conformal time because in these coordinates the FLRW metric is said to be conformally related to metric Minkowski metric. The usual cosmic time $t$ is related to $\tau$ via $dt=ad\tau$.

Now, we need some equations for the dynamics of the universe, which are the Einstein Equations $\tilde G_{\mu\nu}=M_{\rm pl}^{-2} \tilde T_{\mu\nu}$. Let me compute them very quickly by expanding $\tilde G_{\mu\nu}$ in terms of  $a$ and $g_{\mu\nu}$ (this is called a conformal transformation and I left a nice exercise below), which yields
\begin{align}\label{eq:Gmunugeneral}
\tilde G_{\mu\nu}[\tilde g]=G_{\mu\nu}[g]+2\left(g_{\mu\nu}\nabla_\alpha\nabla^\alpha-\nabla_\mu\nabla_\nu\right)\ln a+2\nabla_\mu\ln a\nabla_\nu\ln a+g_{\mu\nu}\nabla_\alpha\ln a\nabla^\alpha\ln a\,.
\end{align}
Looking at the $00$ and $ij$ components and setting $h_{\mu\nu}=0$ we respectively get
\begin{align}
\tilde G_{00}[\tilde g]=3\frac{a'^2}{a^2}=3{\cal H}^2
\quad{\rm and}\quad
\tilde G_{ij}[\tilde g]=-\left(2{\cal H}'+{\cal H}^2\right)\delta_{ij}\,,
\end{align}
where ${\cal H}$ is the conformal Hubble parameter and $\delta_{ij}$ is the Kronecker delta. ${\cal H}$ is related to the standard Hubble parameter, $H$, by ${\cal H}=aH$. Note that Eq.~\eqref{eq:Gmunugeneral} is general and $G_{\mu\nu}[g]=G_{\mu\nu}[\eta+h]$ contains all perturbations. Also note that $G_{\mu\nu}[\eta]=0$. We will come back to this expansion at the end of the lecture to talk about the energy density (or backreaction) of GWs.\\

\begin{tcolorbox}[title=\bf{Try it!},coltitle=black,colback=white,colframe=orange!50]
Use \textsc{Mathematica} package \href{http://xact.es/xCoba/}{\textsc{xCoba}} to compute the Einstein Tensor of the FLRW universe. To use \textsc{xCoba} you will need to first install \href{http://xact.es/}{\textsc{xAct}}. You can use \textsc{xCoba} to compute any tensor components of a given metric in a given set of coordinates.
\end{tcolorbox}

\begin{tcolorbox}[title=\bf{Exercise},colback=white,colframe=black!50]
If you work in gravity and cosmology, you will eventually deal with conformal transformations (=local rescaling of the whole metric). Show that if we define
\begin{align}
\tilde g_{\mu\nu}=\Omega^2(x^\mu)g_{\mu\nu}\,,
\end{align}
then the Christoffel symbols are related by
\begin{align}
\tilde \Gamma^\alpha_{\mu\nu}= \Gamma^\alpha_{\mu\nu}+2\delta^\alpha_{(\mu}\nabla_{\nu)}\Omega-g_{\mu\nu}\nabla^\alpha\ln \Omega\,.
\end{align}
A parenthesis in the indices means normalized symmetrization, i.e. $(a,b)=\tfrac{1}{2}(a+b)$.
If you feel strong, you can also show that
\begin{align}
\tilde R_{\mu\nu}=R_{\mu\nu}-2\nabla_\mu\nabla_\nu\ln\Omega-g_{\mu\nu}\nabla_\alpha\nabla^\alpha\ln\Omega+2\nabla_\mu\ln\Omega\nabla_\nu\ln\Omega-2g_{\mu\nu}\nabla_\alpha\ln\Omega\nabla^\alpha\ln\Omega\,.
\end{align}
Then you can move on and check $\tilde G_{\mu\nu}=\tilde R_{\mu\nu}-\tfrac{1}{2}\tilde g_{\mu\nu}\tilde R$. You may want to check Robert Wald's book on General Relativity, appendix D.
\end{tcolorbox}

Cosmology wouldn’t be cosmology if there were no matter in the universe. Matter fields are taken into account with the energy momentum tensor. Because of the homogeneity and isotropy of the universe we often consider either an adiabatic perfect fluid (without anisotropic stress) or a scalar field (which is also a perfect fluid). For an adiabatic perfect fluid we have
\begin{align}\label{eq:tmunu1}
\tilde T_{\mu\nu}=(\rho+p)\tilde u_\mu\tilde u_\nu+p\tilde g_{\mu\nu}\,,
\end{align}
where $\rho$ is the energy density, $p$ the pressure and $\tilde u_\mu$ is the 4-velocity of the fluid. We say it is an adiabatic perfect fluid when $p=w\rho$ where $w$ is the equation of state of the fluid. For radiation (collection of ultrarelativistic particles) we have $w=1/3$. The 4-velocity is normalized to be time-like and unity, namely $\tilde g^{\mu\nu}\tilde u_\mu\tilde u_\nu=-1$. It is customary to take $\tilde u_\mu=a^{-1}u_\mu$ where $u_\mu=(-1,v_i)$, $v_i$ is the spatial velocity which is small and treated as perturbation. For a scalar field $\varphi$ we instead have
\begin{align}\label{eq:tmunu2}
\tilde T_{\mu\nu}=\partial_\mu\varphi\partial_\nu\varphi-\tilde g_{\mu\nu} \left(\tfrac{1}{2}\partial_\alpha\varphi\partial^\alpha\varphi+V(\varphi)\right)\,.
\end{align}
Here $V(\varphi)$ is a general potential for the scalar field. The scalar field can also be interpreted as a perfect fluid by identifying $\rho_\varphi=\tfrac{1}{2}\dot\varphi+V$ and $p_\varphi=\tfrac{1}{2}\dot\varphi-V$, where $\dot\varphi=d\varphi/dt$.

With all this machinery we can finally write the famous Friedmann equations, which read
\begin{align}\label{eq:firstfriedmann}
3{\cal H}^2M_{\rm pl}^2=a^2\rho\quad{\rm and}\quad \left(2{\cal H}'+{\cal H}^2\right)M_{\rm pl}^2=-a^2p\,.
\end{align}
By taking the time derivative of the first Friedmann equation and using ${\cal H}'$ from the second, we recover the energy conservation equation for the perfect fluid, namely
\begin{align}
\rho'+3{\cal H}(1+w)\rho=0\,.
\end{align}
This is solved readily in terms of $a$ for a constant $w$. The solution is given by
\begin{align}\label{eq:rhoomega}
\rho=\rho_*\left(\frac{a}{a_*}\right)^{-3(1+w)}\,,
\end{align}
where we normalized it to an arbitrary “pivot” time “$_*$”. Note that for the scalar field we in general have $w\neq \rm constant$ and so we have to solve the coupled differential equations. In some cases, like for an exponential potential the equations can be solved analytically \cite{Lucchin:1984yf}. Proceeding with $w={\rm constant}$ and Eq.~\eqref{eq:rhoomega}, we can solve the first Friedmann equation \eqref{eq:firstfriedmann} which gives
\begin{align}\label{eq:generala}
a(\tau)=a_*\left(\frac{\tau}{\tau_*}\right)^{1+b}\quad{\rm with}\quad b=\frac{1-3w}{1+3w}\,.
\end{align}
For a radiation dominated universe ($w=1/3$) we have that $a\sim \tau$, ${\cal H}\sim 1/\tau$ and $\rho\sim a^{-4}$. This is all we need to know from the “background” dynamics of the FLRW universe. \\

\begin{tcolorbox}[title=\bf{Curiosity},colback=white,colframe=greenish!50]
There is an exact solution to a universe filled with radiation ($w=1/3$) and pressureless matter ($w=0$). In that case, we have to solve
\begin{align}
 3{\cal H}^2M_{\rm pl}^2=a_*^2\left[\rho_{\rm rad,*}\left(\frac{a}{a_*}\right)^{-2}+\rho_{\rm mat,*}\left(\frac{a}{a_*}\right)^{-1}\right]\,.
\end{align}
To find the exact solution it is convenient to choose the pivot time $a_*$ as the time when the energy density of radiation and matter are equal, i.e. $\rho_{\rm rad,eq}=\rho_{\rm mat,eq}= \tfrac{3}{2}{\cal H}_{\rm eq}^2M_{\rm pl}^2$ at $a_*=a_{\rm eq}$. Integrate (it's a basic integral) and show that
\begin{align}
\frac{a}{a_{\rm eq}}=2\left(\frac{\tau}{\tau_o}\right)+\left(\frac{\tau}{\tau_o}\right)^{2}\,,
\end{align}
where $(\sqrt{2}-1)\tau_o=\tau_{\rm eq}$. Here I used that $1/\tau_{\rm eq}=\sqrt{{2a_{\rm eq}^2\rho_{\rm eq}}/({3M_{\rm pl}^2})}={\cal H}_{\rm eq}$.
Or check Mukhanov's book “Physical Foundations of Cosmology”. We will use the time of radiation-matter equality several times later. It's quite convenient as the quantities, such as the Hubble radius, at the time of radiation-matter equality are well measured by CMB observations, like the Planck satellite.
\end{tcolorbox}

\subsection{What do we call GWs in cosmology? \label{sub:GWs?}}

What may sound like a stupid question at first, it is actually not that trivial to answer and it is in a strict sense not totally clear to this day. This has nothing to do with the detection of GWs but, rather, on how we connect the predictions of the theory with what we detect (more on this later).

To understand what we call GWs in cosmology, we have to review first the so-called scalar, vector and tensor (a.k.a. helicity) decomposition of perturbations on top of a FLRW background (this also means there is a well defined notion of time and space, and space is homogeneous and isotropic). Simply put we can split our perturbation into separate components, which at linear level decouple (are independent) from each other. You can find some details on the decomposition in App.~\ref{app:decomposition}.

With the scalar-vector-tensor decomposition we have we can write the perturbed flat FLRW metric as
\begin{align}\label{eq:perturbFLRW}
d\tilde s^2=a^2(-(1+2\phi)d\tau^2+B_idx^id\tau+H_{ij}dx^idx^j)\,.
\end{align}
In a more standard notation in cosmology, we can write $B_i=\tilde B_i+\partial_i B$, with $\partial^i\tilde B_i=0$, and $H_{ij}$ as
\begin{align}
H_{ij}=2\psi \delta_{ij}+2(\partial_i\partial_j-\frac{1}{3}\delta_{ij}\Delta) E+\partial_{(i}F_{j)}+h_{ij}\,,
\end{align}
with $\partial_iF^i=0$ and $\delta^{ij}h_{ij}=\partial^ih_{ij}=0$. Vector modes are often unimportant in cosmology as they decay. So we will neglect them. We have four scalars: $\phi$, $B$, $\psi$ and $E$. But only two are independent due to diffeomorphism invariance. This means that we can kill two of them by what is called fixing the gauge. And, we have tensor modes $h_{ij}$, which are independent on the gauge at linear order. Note that I am neglecting the matter sector, which is important, but it will not play a very important role in the discussion besides being a source term to $h_{ij}$ later.

To see the meaning of $h_{ij}$, let's forget about cosmology for a second by looking at tensor modes at very sub-Hubble scales (that is drop $a$ and scalar and vector components). Then the perturbed FLRW metric is
\begin{align}\label{eq:perturbFLRW2}
ds^2\approx -d\tau^2+(\delta_{ij}+h_{ij})dx^idx^j\,,
\end{align}
again with $\delta^{ij}h_{ij}=\partial^ih_{ij}=0$. This is precisely the form the metric for GWs in a Minkowski metric in the so-called transverse-traceless gauge. In summary, all of the above work is to understand that:\\

\noindent\textit{“Tensor modes $h_{ij}$ (at linear order) can be interpreted as GWs when deep inside the Hubble radius.”}\\

\noindent In the opposite limit, tensor modes $h_{ij}$ with wavelengths larger than the Hubble radius are not waves. In fact, they can be interpreted as an almost constant anisotropic stress. And, we can only know about super-Hubble tensor modes mean once they enter the Hubble radius.

\subsection{Behaviour of GWs in cosmology \label{sec:GWsincosmo}}

We have identified the components of the metric which correspond to GWs on sub-Hubble scales, the tensor modes $h_{ij}$. We now need to understand how they behave, especially how they become GWs and how their amplitude changes with the expansion of the universe. 

The equations of motion for $h_{ij}$ follow those of a massless scalar field, namely at linear order
\begin{align}\label{eq:hij}
h_{ij}''+2{\cal H}h'_{ij}-\Delta h_{ij}=0\,.
\end{align}
The right hand side of \eqref{eq:hij} is only non-vanishing at leading order if there is anistropic stress in the universe. But we have not found any departure from isotropy and at linear order scalar, vector and tensor decouple. So at leading order tensor modes are basically free massless fields. Curiously, free streaming particles like neutrinos can have a significant backreaction to GWs by damping their amplitude about $10\%$ \cite{Weinberg:2003ur}. Free streaming particles somehow act like a resistance of a medium to be deformed by GWs, thus GW lose energy.\\

\begin{tcolorbox}[title=\bf{Exercise},colback=white,colframe=black!50]
Show that
\begin{align}
R^{(1)}_{\mu\nu}[\eta+h]=\frac{1}{2}\left(2\partial_\alpha\partial_{(\mu}h_{\nu)}\,^\alpha-\partial_\alpha\partial^\alpha h_{\mu\nu}-\partial_\mu\partial_\nu h\right)\,,
\end{align}
where $h=\eta^{\mu\nu}h_{\mu\nu}$, and recover the equations of motion for the transverse-traceless component of the spatial metric of the perturbed FLRW universe.
\end{tcolorbox}

Now, tensor modes have two polarizations, probably the most well-known ones are the plus and cross polarizations. We can capture this in the Fourier expansion with the polarization tensors $\epsilon^\lambda_{ij}$, where $\lambda$ is the polarization. Namely, the simplest Fourier expansion we can take is given by
\begin{align}\label{eq:hijfourier}
h_{ij}=\frac{1}{(2\pi)^3}\sum_\lambda\int d^3k \,h_{\bm{k},\lambda}(\tau)\epsilon_{ij}^\lambda(\bm{k})\, e^{i\bm{k}\cdot\bm{x}}\,.
\end{align}
The polarization tensors are orthogonal to the direction of propagation $k^i\epsilon_{ij}^\lambda(\bm{k})=0$ and they are traceless $\delta^{ij}\epsilon_{ij}^\lambda(\bm{k})=0$. You can find the explicit expressions of such tensors along a general direction $k$ in polar coordinates in App.~\ref{app:polarization}. In the form \eqref{eq:hijfourier}, the reality condition of $h_{ij}$ translates into $h^*_{\bm{k},\lambda}=h_{-\bm{k},\lambda}$ and $(\epsilon_{ij}^\lambda(\bm{k}))^*=\epsilon_{ij}^\lambda(-\bm{k})$. The normalization of the polarization tensors is
\begin{align}
(\epsilon_{ij}^\lambda(\bm{k}))^*\epsilon^{ij,\lambda'}(\bm{k})=\delta^{\lambda\lambda'}\,.
\end{align}

Note that you can check using the formulas in appendix \ref{app:polarization} that the reality condition $(\epsilon_{ij}^\lambda(\bm{k}))^*=\epsilon_{ij}^\lambda(-\bm{k})$ is not satisfied by the plus and cross (especially the cross) polarization. If we want to use plus and cross we could have included the complex conjugate directly in Eq.~\eqref{eq:hijfourier}, just like when we expand a field into creation and annihilation operators in quantum field theory. For simplicity, I will insist with Eq.~\eqref{eq:hijfourier} and instead use Right and Left (circular) polarizations. They are related with the plus and cross polarizations by
\begin{align}
e^{R}_{ij}(\mathbf{k})=\tfrac{1}{\sqrt{2}}(e^{+}_{ij}(\mathbf{k})+ie^{\times}_{ij}(\mathbf{k}))\quad{\rm and}\quad e^{L}_{ij}(\mathbf{k})=\tfrac{1}{\sqrt{2}}(e^{+}_{ij}(\mathbf{k})-ie^{\times}_{ij}(\mathbf{k}))\,.
\end{align}
\begin{tcolorbox}[title=\bf{Try it!},coltitle=black,colback=white,colframe=orange!50]
If you want to picture how a GW with circular polarization looks like, I suggest you have a look at the animation at \href{https://www.einstein-online.info/en/spotlight/gw_waves/}{Einstein-online}.
\end{tcolorbox}
The equations of motion \eqref{eq:hij} for the tensor modes for a given polarization then read
\begin{align}\label{eq:tensormodeq}
h_{\bm{k},\lambda}''+2{\cal H}h_{\bm{k},\lambda}'+k^2 h_{\bm{k},\lambda}=0\,.
\end{align}
We can find solutions to equation in the two relevant limits: super-Hubble scales ($k\ll {\cal H}$) and sub-Hubble scales ($k\gg {\cal H}$). On super-Hubble scales we neglect $k$ and have that
\begin{align}\label{eq:superHphi}
h_{\bm{k},\lambda}(k\ll {\cal H})\approx C_1+C_2\int \frac{d\tau}{a^2}\,.
\end{align}
The term proportional to $C_2$ is the so-called decaying mode, which blows up when $a\to 0$. So we set $C_2=0$ if initial conditions are provided by inflation. $C_1$ could be related to the fluctuations at the end inflation. On sub-Hubble scales, we do what is called a \href{https://en.wikipedia.org/wiki/WKB_approximation}{WKB approximation} and select only the oscillating part with the ansatz $h_{\bm{k},\lambda}\sim A e^{\pm ik\tau}$ with $k\gg {\cal H}$, then we solve for $A$. By doing so, we arrive at
\begin{align}\label{eq:subHphi}
h_{\bm{k},\lambda}(k\gg {\cal H})\approx\frac{1}{a}\left( C_3{\rm e}^{i k\tau}+C_4{\rm e}^{-ik\tau}\right)\,.
\end{align}
You can readily see that the amplitude goes as $1/a$ if you change variables to $h_{\bm{k},\lambda}=v_{\bm{k},\lambda}/a$. You will find that the friction term disappears and for $k\gg {\cal H}$ we have $v_{\bm{k},\lambda}''+k^2v_{\bm{k},\lambda}\approx 0$. We can find the values of $C_3$ and $C_4$ by matching at the time when $k={\cal H}$, called the time of Hubble radius crossing, say $\tau_{\cal H}$. This gives $C_4=-C_3$ and $C_3\sim -iC_1 a(\tau_{\cal H})$ so that
\begin{align}\label{eq:superHh}
h_{\bm{k},\lambda}(k\gg {\cal H})\approx C_1\frac{a(\tau_{\cal H})}{a(\tau)}\sin k\tau\,.
\end{align}
This means that, after a given tensor mode enters the Hubble radius, it undergoes oscillations. Looking back at the Fourier expansion for $h_{ij}$ \eqref{eq:hijfourier} we see that when $h_{ij}$ is inside the horizon we have
\begin{align}
h_{ij}\sim {a^{-1}} \sum_\lambda\int d^3k \,\epsilon_{ij}^\lambda(\bm{k})\, e^{\pm ik\tau+i\bm{k}\cdot\bm{x}}\,.
\end{align}
This means that $h_{ij}$ is actually a collection of plane \textit{waves} whose amplitude decays with the scale factor. In this limit, the linear $h_{ij}$ are gravitational waves, as we argued before, and something that we may detect at, e.g., GW interferometers.
\\
\begin{tcolorbox}[title=\bf{Try it!},coltitle=black,colback=white,colframe=orange!50]
Use \textsc{Mathematica} package \href{http://www2.iap.fr/users/pitrou/xpand.htm}{\textsc{xPand}} to check the linear equations for $h_{ij}$. You will need to first install \href{http://xact.es/}{\textsc{xAct}}.
\end{tcolorbox}

\begin{tcolorbox}[title=\bf{Supplementary information},colback=white,colframe=mulberry!50,coltitle=black]
There is an exact solution for the tensor modes $h_{\bm{k},\lambda}$ when the universe has a constant equation of state like in Eq.~\eqref{eq:generala}. In that case, the tensor mode equation \eqref{eq:tensormodeq} reads 
\begin{align}\label{eq:deltaphi2}
h_{\bm{k},\lambda}''+\frac{2(1+b)}{\tau}h_{\bm{k},\lambda}'+ k^2h_{\bm{k},\lambda}=0\,.
\end{align}
The general solution is given in terms of Bessel functions $J_\alpha$ and $Y_\alpha$, namely
\begin{align}\label{eq:generalh}
h_{\bm{k},\lambda}=(k\tau)^{-1/2-b}\left(C_1J_{1/2+b}(k\tau)+C_2Y_{1/2+b}(k\tau)\right)\,.
\end{align}
You can check that $C_1$ corresponds to the “growing” mode which is constant on super-Hubble scales and that you recover the limits we derived above. If initial conditions are set by inflation, then we have that
\begin{align}
C_1=2^{1/2+b}\,\Gamma[b+3/2]\,h^{\rm prim}_{\bm{k},\lambda}\,,
\end{align}
where $\Gamma[x]$ is the Gamma function and $h^{\rm prim}_{\bm{k},\lambda}$ is the primordial fluctuations for the tensor modes generated during inflation.
\end{tcolorbox}

Before we move on to the next point of the lecture, it is important to understand how a GW detector will see GWs that were generated in the very early universe. Consider that a GW was generated at a time $\tau_*$ from a source with size $x_0$ today. In cosmology, sources are mostly isotropically distributed and have a limit on their correlation length (e.g. only those GW within a Hubble are in causal contact and so correlated). Thus, we will take that GWs are emitted with the typical correlation length (“size”) of the source is proportional to the Hubble radius at that time, say $x_*\sim H_*^{-1}$. Then, GWs travelled until they hit the detector for a comoving distance $r$. Since they are massless fields they propagate along null geodesics, i.e. $ds^2=0$ or $dr=d\tau$ (assuming flat FLRW). The physical distance they travelled is then $R=a_0r=a_0(\tau_0-\tau_*)\approx a_0\tau_0\sim H_{0}^{-1}$. Using some basic trigonometry we find
\begin{align}
\sin\theta\sim \theta\sim \frac{x_0}{R}\approx \frac{H_*^{-1}}{H_{0}^{-1}}\frac{a_0}{a_*}=\frac{{\cal H}_0}{{\cal H}_*}\,,
\end{align}
where we used that $x_*=x_0(a_*/a_0)$ and $\theta\ll 1$. So the observed angular size of the “source” is proportional to the ratio of comoving Hubble radius. Juggling a bit with the factors we get
\begin{align}
\theta\sim\frac{H_0a_0}{H_*a_*}\approx \frac{H_0}{H_{\rm eq}}(1+z_{\rm eq})\frac{a_*}{a_{\rm eq}}\approx \frac{H_0}{k_{\rm eq}}\frac{T_{\rm eq}}{T_{*}}
\end{align}
where I used that during radiation domination $H\sim {H_{\rm eq}} (a/a_{\rm eq})^{-2}$, that $1+z=a_0/a$ is the redshift, ${H_{\rm eq}}=k_{\rm eq}(1+z_{\rm eq})$ and that roughly $T\sim 1/a$ is the temperature of the radiation fluid. We neglected any change in the number of relativistic particles, but we will do a bit better than that later. Putting some number from Tabs.~\ref{tab:conversion} and \ref{tab:quantities} we get
\begin{align}
\theta\sim 10^{-15}\left(\frac{T_*}{10^4\,{\rm GeV}}\right)^{-1}\,.
\end{align}
The problem is that GW detectors have a very poor angular resolution, compared say to photon detectors. An optimistic angular resolution for GW detectors is roughly of the order $\theta\sim 10\,{\rm deg}$ \cite{Caprini:2018mtu}. I think it's quite obvious that cosmological sources cannot be resolved. They may still be detectable though as correlated noise in GW detectors though. This is called the stochastic gravitational wave background.

\subsection{Energy density of GWs\label{sub:GWsenergy}}

Since GWs coming from the early universe cannot be individually resolved, we have to turn to correlated noise in the detectors. For example, we can correlate GW signals in two detectors, something like $\langle h_{ij}h^{ij}\rangle$, and look for common noises. The brackets $\langle...\rangle$ denote some kind of averaging procedure. For a GW detector it may mean volume or time average. For us it will mainly mean ensemble average. Luckily, if we have a large enough sample (i.e. the wavelength of the wave is such that we can divide the total volume into many smaller ones and still take a meaning full average) then by the ergodic hypothesis, volume and ensemble average are essentially the same. The actual connection to the GW detectors such as LISA, will be discussed in other lectures of the school. 

Here I would like to focus on the connection with our theoretical predictions. We can compute the two point correlation $\langle h_{ij}h^{ij}\rangle$ in terms of the Fourier expansion \eqref{eq:hijfourier}, namely
\begin{align}
\langle h_{ij}(\bm{x},\tau)h^{ij}(\bm{x}',\tau)\rangle=\frac{1}{(2\pi)^6}\sum_{\lambda,\lambda'}\int d^3k\,d^3k'\,\langle h_{\bm{k},\lambda}(\tau)h_{\bm{k}',\lambda'}(\tau)\rangle\epsilon_{ij}^\lambda(\bm{k}) \,\epsilon^{ij}_{\lambda'}(\bm{k}')\, e^{i\bm{k}\cdot\bm{x}+i\bm{k}'\cdot\bm{x}'}\,.
\end{align}
Now, let us assume that $h_{\bm{k},\lambda}$ is drawn from a random distribution. In the cosmological set up, since the initial conditions are stochastic, we will have a given probability that $h_{\bm{k},\lambda}$ in a Hubble patch has a certain amplitude. Since the sources are basically homogeneously and isotropically distributed, we have that
\begin{align}\label{eq:spectrumh}
\langle h_{\bm{k},\lambda}(\tau)h_{\bm{k}',\lambda'}(\tau)\rangle=(2\pi)^3 \delta(\bm{k}+\bm{k}')\delta_{\lambda\lambda'}P_{h,\lambda}(k,\tau)\,.
\end{align}
Homogeneity leads to the Dirac delta $\delta(\bm{k}+\bm{k}')$, since spatial translations are related to momentum conservation. Isotropy means that the power of fluctuations $P_{h,\lambda}(k)$ only depends on the modulus $k=|\bm{k}|$ not the direction. If you want to read more these brackets in the context of inflation, I recommend \href{https://nms.kcl.ac.uk/eugene.lim/AdvCos/lecture2.pdf}{Eugene Lim's} notes. $P_{h,\lambda}(k)$ is what we can actually compute from a given theory and model. For the two point function, we have
\begin{align}
\langle h_{ij}(\bm{x},\tau)h^{ij}(\bm{x}',\tau)\rangle=\frac{1}{(2\pi)^3}\sum_{\lambda}\int d^3k\,P_{h,\lambda}(k,\tau)\, e^{i\bm{k}(\bm{x}-\bm{x}')}\,.
\end{align}
If we evaluate the two point function at the same point $\bm{x}=\bm{x}'$ then
\begin{align}\label{eq:hijhijdimension}
\langle h_{ij}(\bm{x},\tau)h^{ij}(\bm{x},\tau)\rangle=\sum_{\lambda}\int d\ln k \frac{k^3}{2\pi^2}\,P_{h,\lambda}(k,\tau)=\sum_{\lambda}\int d\ln k\,{\cal P}_{h,\lambda}(k,\tau)\,,
\end{align}
where we introduced the dimensionless power spectrum ${\cal P}_{h,\lambda}(k,\tau)=\frac{k^3}{2\pi^2}\,{ P}_{h,\lambda}(k,\tau)$. This is what we need to know about the two point correlation function.

To understand the connection of this “noise” in the detector and cosmology, we need to understand the effect of GWs on the cosmology itself. That is, the GW backreaction to the background FLRW metric. We can compute this as follows. First, we expand the Einstein Equations in terms of perturbations. I will use the conformal expansion \eqref{eq:Gmunugeneral} so that we can easily neglect the terms due to expansion of the universe. We are only interested in the effect sub-horizon tensor modes. The perturbation expansion leads to
\begin{align}\label{eq:expansion2}
G_{\mu\nu}[\eta+h]=G_{\mu\nu}^{(0)}[\eta+h]+G_{\mu\nu}^{(1)}[\eta+h]+G^{(2)}_{\mu\nu}[\eta+h]\,.
\end{align}
The first piece vanishes $G^{(0)}_{\mu\nu}[\eta+h]$ because $\eta_{\mu\nu}$ is flat Minkowski. The second piece $G^{(1)}_{\mu\nu}[\eta+h]$ gives the linear equations of motion and it vanishes if we take any kind of average, i.e. $\langle G_{\mu\nu}[\eta+h]^{(1)}\rangle=0$, because it is linear in perturbations and they average to zero. The third piece $G^{(2)}_{\mu\nu}[\eta+h]$ is quadratic in $h_{\mu\nu}$ and it does not vanish after taking the average. We can think of this last piece as the backreaction of the fluctuations onto the background. We can also associate to it a (pseudo)-energy momentum tensor as follows. Coming back to the full equation \eqref{eq:Gmunugeneral} we can move  $G^{(2)}_{\mu\nu}[\eta+h]$ to the right hand side and write
\begin{align}
\tilde G_{\mu\nu}^{(0)}[\tilde g]\approx M_{pl}^{-2}\tilde T_{\mu\nu}-\langle G^{(2)}_{\mu\nu}[\eta+h]\rangle\approx M_{pl}^{-2}\left( \tilde T_{\mu\nu}+ t^{\rm GW}_{\mu\nu}\right)\,,
\end{align}
where
\begin{equation}\label{eq:pseudotensor}
t_{\mu\nu}^{\rm GW}=\frac{M_{pl}^2}{4}\bigg\langle\partial_\mu h^{\alpha\beta}\partial_\nu h_{\alpha\beta}-\frac{1}{2}\tilde g_{\mu\nu}\tilde \partial_\sigma h^{\alpha\beta}\tilde \partial^\sigma h_{\alpha\beta}\bigg\rangle\,.
\end{equation}
This is called the Isaacson prescription, presented in 1967 by Isaacson in \cite{Isaacson:1967zz}. \\

\begin{tcolorbox}[title=\bf{Exercise},colback=white,colframe=black!50]
Show that
\begin{align}
R^{(2)}_{\mu\nu}[\eta+h]=&\frac{1}{4}\partial_\mu\partial_\nu\left(h^{\alpha\lambda}h_{\alpha\lambda}\right)-\partial_\alpha\left(h^{\alpha\lambda}\Gamma^{(1)}_{\lambda\mu\nu}\right)+\frac{1}{2}\partial_\alpha h \,\eta^{\alpha\lambda}\Gamma^{(1)}_{\lambda\mu\nu}\nonumber\\&
-\frac{1}{4}\partial_\mu h_{\alpha\lambda}\partial_\nu h^{\alpha\lambda}+\frac{1}{2}\left(\eta^{\alpha\sigma}\eta^{\beta\lambda}-\eta^{\alpha\beta}\eta^{\sigma\lambda}\right)\partial_\alpha h_{\lambda(\mu}\partial_\sigma h_{\nu)\beta}\,,
\end{align}
where 
\begin{align}
\Gamma^{(1)}_{\lambda\alpha\beta}[h]=\partial_{(\alpha} h_{\beta)\lambda}-\tfrac{1}{2}\partial_{\lambda} h_{\alpha\beta}\,,
\end{align}
and arrive at the pseudo tensor \eqref{eq:pseudotensor} focusing only on the spatially transverse-traceless degress of freedom in $h_{\mu\nu}$. 
\end{tcolorbox}

Let us briefly stop and understand how we computed $t_{\mu\nu}^{\rm GW}$ and what it means. We are considering GWs with a wavelength much smaller than the Hubble radius (i.e. $\lambda\gg H^{-1}$), which means that in a given volume (smaller than the Hubble volume so as to make sense) we can fit many GW wavelengths. Thus, the brackets in the Isaacson prescription mean average within a volume large and small enough (e.g. its radius sufficiently larger than $\lambda$ but sufficiently smaller than $H^{-1}$). For the frequencies accessible to GW detectors (very small $\lambda$ compared to the size of the Universe), it is not so difficult to find such condition. Now, we can connect the energy density with the noise measured at a GW detector through the average $\langle h_{ij}h^{ij}\rangle$. The noise that is measured in the detector is related to the energy density contained in the GW background.

By analogy with the energy momentum tensor of a perfect fluid \eqref{eq:tmunu1}, we find the energy density of GWs by computing the $00$ component of $t_{\mu\nu}^{\rm GW}$, namely (working in conformal coordinates)
\begin{align}
a^2\rho_{\rm GW}=t_{00}^{\rm GW}=\frac{M_{pl}^2}{8}\bigg\langle h'^{ij} h'_{ij}+ \partial_k h^{ij} \partial^k h_{ij}\bigg\rangle\,.
\end{align}
Going to Fourier space we have
\begin{equation}\label{eq:spectumdensity}
\rho_{\rm GW}=\frac{M_{\rm pl}^2}{8a^2}\int d\ln k \frac{k^3}{2\pi^2}\left\{P_{h',\lambda}(k,\tau)+ k^2 P_{h,\lambda}(k,\tau)\right\}\,,
\end{equation}
where we defined $P_{h',\lambda}(k,\tau)$ through $\langle h'_{\bm{k},\lambda}h'_{\bm{k}',\lambda}\rangle$. To simplify calculations, we can use that since $h_{\bm{k},\lambda}$ is like plane wave, we have $h'_{\bm{k},\lambda}\approx k h_{\bm{k},\lambda}$ where we can safely neglect the expansion of the universe. Note that since $h\sim 1/a$ inside the Hubble radius, we have that $\rho_{\rm GW}\sim 1/a^4$ and, therefore, GWs behave as a radiation fluid.

Finally, we can compute what is called the spectral density, which is given by
\begin{equation}\label{eq:spectraldensity}
\Omega_{\rm GW}(k,\tau)\equiv\frac{1}{3 M_{\rm pl}^2 H^2}\frac{d \rho_{\rm GW}}{d\ln k}=\frac{k^2}{12a^2H^2}\sum_\lambda{\cal P}_{h,\lambda}(k)\,,
\end{equation}
and describes the fraction of energy density in GWs in the universe  per logarithmic wavenumber. We only have one thing left to do. Estimate the amplitude of such spectral density if GW were generated in the very early universe. Note that during radiation domination $\Omega_{\rm GW}(k,\tau)$ is basically constant.

Let's say that we generated some GWs at a time $t_*$, let me call their spectral density at that time $\Omega_{\rm GW,*}$. What is the amplitude of the GW spectral density that we measure today, say $\Omega_{\rm GW,0}$? There is some nice way to compute it. We first multiply $\Omega_{\rm GW,0}$ by $h^2=H_0^2/H_{100}^2$ where $H_{100}=100{\rm km/s/Mpc}$ (this is to take into account the uncertainty in $H_0$) and then we juggle with the factors, like this:
\begin{align}
\Omega_{\rm GW,0}h^2&=\frac{1}{3H_{100}^2M_{\rm pl}^2}\frac{d\rho_{\rm GW,0}}{d\ln k}=\frac{\rho_{\rm rad,0}}{3H_{100}^2M_{\rm pl}^2}\frac{1}{\rho_{\rm rad,0}}\frac{d\rho_{\rm GW,0}}{d\ln k}\nonumber\\&=\Omega_{\rm rad,0}h^2\frac{1}{\rho_{\rm rad,0}}\frac{d\rho_{\rm GW,0}}{d\ln k}=\Omega_{\rm rad,0}h^2\frac{\rho_{\rm rad,*}}{\rho_{\rm rad,0}}\left(\frac{a_0}{a_*}\right)^{-4}\Omega_{\rm GW,*}\,.
\end{align}
We are left to compute the factor involving $\rho_{\rm rad}$ and $a$, which actually would be unity if not were for the fact that the number of relativistic species (radiation) changes with the temperature. Taking into account the change (see App.~\ref{app:formulasuseful}) we have
\begin{align}
\frac{\rho_{\rm rad,*}}{\rho_{\rm rad,0}}\left(\frac{a_0}{a_*}\right)^{-4}=\frac{g_{\rho}(T_*)T_*^4}{g_{\rho}(T_0)T_0^4}\times\left(\frac{T_*}{T_0}\left(\frac{g_{s}(T_*)}{g_{s}(T_0)}\right)^{1/3}\right)^{-4}=\frac{g_{\rho}(T_*)}{g_{\rho}(T_0)}\left(\frac{g_{s}(T_0)}{g_{s}(T_*)}\right)^{4/3}\,.
\end{align}
Using the numbers in Tab.~\ref{tab:quantities} we find that 
\begin{align}\label{eq:GWstoday}
\Omega_{\rm GW,0}h^2&=1.62\times 10^{-5}\left(\frac{\Omega_{\rm rad,0}h^2}{4.18\times 10^{-5}}\right)\left(\frac{g_{\rho}(T_*)}{106.75}\right)\left(\frac{g_{s}(T_*)}{106.75}\right)^{-4/3}\Omega_{\rm GW,*}\,,
\end{align}
where we took that in the standard model of particle physics ${g_{\rho}(T_*)}={g_{s}(T_*)}={106.75}$ for $T_*\gg 100\,{\rm GeV}$. In the next lecture, we will compute $\Omega_{\rm GW,*}$ for various sources related to PBHs.\\

\begin{tcolorbox}[title=\bf{Exercise},colback=white,colframe=black!50]
We can do a similar exercise for the frequency. Imagine that GWs were generated at the time $t_*$ with a typical frequency $k_*={\cal H}_*$. Check that the observed frequency today, assuming a radiation dominated universe, is given by
\begin{align}\label{eq:f0s}
f_{*,0}=\frac{k_*}{2\pi a_0}=2\times 10^{-3}{\rm Hz}\,\left(\frac{T_*}{5\times 10^4\,{\rm GeV}}\right)\left(\frac{g_\rho(T_{*})}{106.75}\right)^{1/2}\left(\frac{g_{s}(T_{*})}{106.75}\right)^{-1/3}\,.
\end{align}
GWs generated when the universe when the universe was roughly ${T_*}\approx{10^{4}\sim10^{5}\,{\rm GeV}}$ will fall inside the peak sensitivity of LISA.
\end{tcolorbox}

\subsubsection{Subtleties in the definition of GW energy density}

I have repeated a couple of times that GWs (more appropriately tensor modes) must have a wavelength smaller than the Hubble radius to make sense of them as GWs. There are a couple or three main caveats in cosmology:
\begin{enumerate}
\item When exactly do tensor modes start behaving as GWs? From which moment on we can use the Isaacson prescription? This is not very clear.
\item Do tensor modes that are super-Hubble backreact onto the background metric? Can we still trust the Isaacson prescription? I think it is clear that the effect of super-Hubble modes should become less and less important as the become more and more super-Hubble. There is some research trying to build a proper energy-momentum tensor.
\item In cosmology we not only have tensor modes; we have scalar and vector. Plus, general relativity is quite non-linear, so we have mixing effects. For example, if $\varphi$ is a scalar, the quantity $\partial_i\varphi\partial_j\varphi$ has a non-vanishing transverse-traceless component. For the same reason, the notion of transverse-traceless degrees of freedom, the tensor modes, changes if we do a coordinate (gauge) transformation up to second order in perturbation theory. In the most strict sense, the Isaacson prescription is dependent on the spacetime slicing. \\
\end{enumerate}

\begin{tcolorbox}[title=\bf{Exercise for the brave},colback=white,colframe=red!70, coltitle=white]
Do a time reparametrisation, e.g. $\bar\tau=\tau+T$, and compute the transformation for tensor modes at second order in $T$. For the metric, you can use the ansatz \eqref{eq:perturbFLRW} but it is more elegant (and might help you if you ever work with second order perturbation theory) to use
\begin{align}
d\tilde s^2=a^2\left(-e^{2\phi}d\tau^2+B_idx^id\tau+e^{2\psi}\left(e^Y\right)_{ij}dx^idx^j\right)\,,
\end{align}
where $\left(e^Y\right)_{ij}$ has to be understood as an exponential matrix, that is 
\begin{align}
\left(e^Y\right)_{ij}=\delta_{ij}+Y_{ij}+\frac{1}{2}Y_{ik}Y^k_j+...\,.
\end{align}
Furthermore, by construction $\det\left(e^Y\right)_{ij}=1$ and which means that $\frac{d}{d\tau}\det\left(e^Y\right)_{ij}=\left(e^{-Y}\right)^{ij}\frac{d}{d\tau}\left(e^Y\right)_{ij}=0$. We can then take
\begin{align}
Y_{ij}=2(\partial_i\partial_j-\tfrac{1}{3}\delta_{ij}\Delta){\cal E}+\partial_iF_j+h_{ij}\,.
\end{align}
Forget all scalar and vector quantities and show that
\begin{align}\label{eq:gaugetransformationhij}
\bar h_{ij}=h_{ij}+\widehat{TT}^{kl}_{ij}\left[T \,h'_{kl}+\partial_kT\partial_l T\right]
\end{align}
where $\widehat{TT}^{kl}_{ij}$ is the transverse-traceless projector, see App.~\ref{app:decomposition}. I provide some useful formulas about gauge transformations in App.~\ref{app:gaugetransformations}. An alternative approach using the Hamiltonian can be found in \cite{Domenech:2017ems}. What Eq.~\eqref{eq:gaugetransformationhij} tells us is that what we call tensor modes in one “gauge” is not fully what we would call tensor modes in another “gauge”, at least at second order. Most importantly, if $\bar\rho_{\rm GW}\sim \langle \bar h_{ij}\bar h^{ij}\rangle$ then it is obvious that $\bar\rho_{\rm  GW}\neq\rho_{\rm  GW}$. This issue is more or less okay for linear tensor modes but it gets troublesome for secondary GWs which we will study in the next lecture.
\end{tcolorbox}
\newpage

\chapter*{Lecture 2}
\addcontentsline{toc}{chapter}{Lecture 2}
\markboth{Lecture 2}{Lecture 2}

\section{Collection of GW signals related to PBHs \label{sec:3}}

Most of the lectures of the school focus on how to generate large fluctuations during inflation, how PBHs form and how to compute their abundance. But, more interesting to me, is that PBH formation requires large fluctuations and large fluctuations mean a large GW signal (by second order effects). In addition to that, PBHs themselves may source GWs by, e.g., mergers, hyperbolic encounter and even by Hawking evaporation. There is also the possibility that whatever amplified fluctuations during inflation also enhanced tensor fluctuations. This latter possibility is more model dependent and so I will not consider it here.

Let us for simplicity assume that PBHs are formed by the collapse of primordial fluctuations (although it is not the only possibility) with a very peaked primordial spectrum, say at $k_{\rm f}$. From here and on, the subscript “f” means PBH formation. So $k_{\rm f}$ is the mode that leads to PBH formation at a time $k_f={\cal H}_{\rm f}$. This implies that PBHs all have one typical mass, call it $M_{\rm PBH,f}$, related to $k_{\rm f}$ by
\begin{align}\label{eq:mpbhf}
 M_{\rm PBH,f}=4\pi\gamma\frac{M_{\rm pl}^2}{H_{\rm f}}\,.
\end{align}
I will also assume that you are familiar with this formula from the other lectures. I will take the standard value of $\gamma=0.2$ otherwise stated. It is important to keep the subscript “f” because PBHs lose mass by evaporation. The monochromatic mass assumption is not very realistic as one rather expects some not so sharp log-normal-like distribution. But we need it to do the calculations analytically. Because we have a typical mass and all signals are related to these PBHs, the typical frequency of the GWs will be related to $M_{\rm PBH,f}$. How proportional? We will give more accurate estimates later but, for now, let us split two cases:\footnote{Initially the two scales we consider are the same but once a PBH forms it decouples from the expansion. Thus, at late times we can split the GWs from physics around the BH radius and physics at the Hubble radius.}
\begin{enumerate}
\item \textbf{GWs generated by local BH physics:} these include PBHs mergers and GWs from Hawking evaporation. In both cases, the typical scale is the size of the PBH (remember the Schwarzschild radius $r_{\rm PBH}=2GM_{\rm PBH,f}$). We will also assume that these GWs are emitted at a time $a_{\rm emit}$. So, as an order of magnitude estimate we have that the generated GWs are observed today  with a frequency given by
\begin{align}\label{eq:frpbh}
f_{r_{\rm PBH},0}=\frac{a_{\rm em}}{a_0}f_{r_{\rm PBH}}= \frac{a_{\rm em}}{a_0}\frac{1}{2\pi r_{\rm PBH}}= \frac{1}{1+z_{\rm em}}\frac{2M_{\rm pl}^2}{M_{\rm PBH,f}}= \frac{2\times 10^{37}\,{\rm Hz}}{1+z_{\rm em}} \left(\frac{M_{\rm PBH,f}}{1\,{\rm g}}\right)^{-1}\,,
\end{align}
where the $2\pi$ comes from the relation between orbital frequency and frequency $f=\omega/2(\pi)$ (or the relation between frequency and wavenumber $f=k/(2\pi)$).
This is an absurdly high frequency signal for light PBHs, even if we take into account that for very light PBHs such that they evaporated in the early universe the redshift of emission is $z_{\rm em}\gg1$. But, for heavy PBHs like $M_{\rm PBH,f}\sim O(M_\odot)$ we have $f_{\rm PBH}\sim O(10^4)\,{\rm Hz}$ assuming they are merging nearby so $z_{\rm em}\ll 1$.

\item \textbf{GWs generated by “large” scale physics} (large compared to the size of the PBHs): these are GWs from primordial fluctuations and GWs from PBH number density fluctuations. Here we have two typical scales: the horizon scale at PBH formation, let us call it ${{\cal H}_{\rm f}}$, and the mean inter-PBH separation scale, say $d_{\rm PBH,f}\sim (4\pi n_{\rm PBH, f}/3)^{-1/3}$. I used the notation $n_{\rm PBH, f}$ as the number density of PBHs at formation. As rough order of magnitude estimates we respectively have
\begin{align}\label{eq:fhpbh}
f_{{\cal H}_{\rm f},0}= \frac{{\cal H}_{\rm f}}{2\pi a_0}= \frac{a_{\rm f}}{a_0}\frac{2\gamma M_{\rm pl}^2}{M_{\rm PBH,f}}=  \frac{2\gamma }{1+z_{\rm f}}f_{r_{\rm PBH}}
\end{align}
and
\begin{align}\label{eq:interfhpbh}
f_{\rm inter-BH,0}= \frac{a_{\rm f}}{a_0}\frac{1}{2\pi d_{\rm PBH, f}}=\frac{a_{\rm f}}{2\pi a_0 }\left(\frac{4\pi}{3}n_{\rm PBH, f}\right)^{1/3}= \frac{\beta^{1/3}}{\gamma^{1/3}}f_{{\cal H}_{\rm f},0}= \frac{2\gamma^{2/3}}{1+z_{\rm f}}{\beta^{1/3}}f_{r_{\rm PBH}}\,,
\end{align}
Where in the last equation I used that $\rho_{\rm PBH, f}=n_{\rm PBH, f}\times M_{\rm PBH, f}$ and $\beta=\rho_{\rm PBH, f}/(3H_f^2M_{\rm pl}^2)$, which measures the fraction of energy density in the form of PBHs at formation.
\end{enumerate}

\begin{figure}
\centering
\includegraphics[width=0.6\columnwidth]{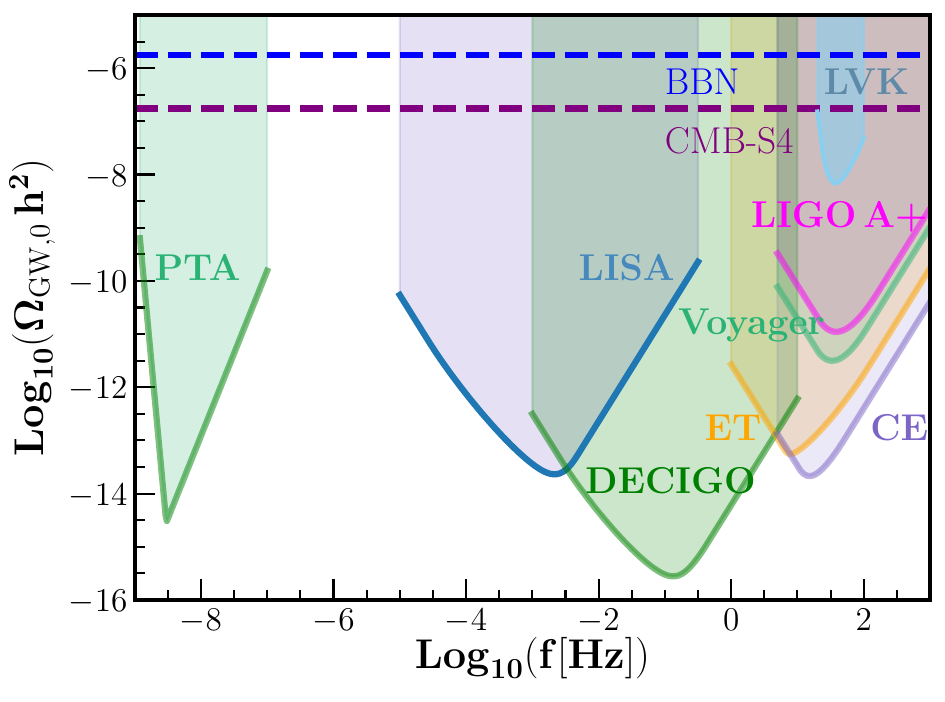}
\caption{Power-law integrated sensitivity curves~\cite{Thrane:2013oya} for PTA, LISA, DECIGO, Einstein Telescope (ET), Cosmic Explorer (CE), Voyager and LIGO A+ experiments. You can find the sensitivity curves in \cite{ce,A+,voyager,Schmitz:2020syl}. In light blue you can see the current upped bounds bounds on the GW background from the LIGO/Virgo/KAGRA collaboration \cite{KAGRA:2021kbb}. The horizontal thick long dashed blue and purple lines respectively show the current constraint from BBN \cite{Cyburt:2004yc,Arbey:2021ysg,Grohs:2023voo} (in blue) and from future CMB-S4 experiments \cite{CMB-S4:2016ple,Arbey:2021ysg}.\label{fig:sensitivity curves}}
\end{figure}

\newpage

\begin{tcolorbox}[title=\bf{Exercise},colback=white,colframe=black!50]
To understand the difference between cases $1$ and $2$ compute the factor $1+z_{\rm f}$. To do things slightly more complicated, assume that there was a phase of matter domination ($w=0$) before reheating the universe at $t_{\rm rh}$ with temperature $T_{\rm rh}$. The answer is
\begin{align}
1+z_f\approx2.5\times 10^{32}\left(\frac{M_{\rm PBH,f}}{1\,\rm g}\right)^{-2/3}\left(\frac{T_{\rm rh}}{5\times 10^4\,{\rm GeV}}\right)^{-1/3}\left(\frac{g_s(T_{\rm rh})}{g_\rho(T_{\rm rh})}\right)^{1/3}\,,
\end{align}
where I used \eqref{eq:firstfriedmann}, \eqref{eq:rhoomega}, \eqref{eq:mpbhf}, \eqref{eq:HofT} and $\gamma=0.2$. In the limit $T_{\rm rh}\to T_{\rm f}$ (no matter domination at all), check that 
\begin{align}\label{eq:zf}
1+z_f\approx5.5\times 10^{28}\left(\frac{M_{\rm PBH,f}}{1\,\rm g}\right)^{-1/2}\left(\frac{g_\rho(T_{\rm f})}{106.75}\right)^{-1/4}\left(\frac{g_s(T_{\rm f})}{106.75}\right)^{1/3}\,.
\end{align}
You can convince yourself using \eqref{eq:fhpbh} that the frequency of GWs generated by large scale physics can enter the frequency range of GW detectors for heavy enough PBHs (see Fig.~\ref{fig:sensitivity curves}). 
\end{tcolorbox}

\begin{tcolorbox}[title=\bf{Supplementary information},colback=white,colframe=mulberry!50,coltitle=black]
Since GWs contribute to the energy density in the universe, there are other ways to indirectly detect their effect. Any additional radiation component would affect the predictions of Big Bang Nucleosynthesis (BBN) and the CMB. The time relevant for BBN is somewhere around $T\sim 0.1\,{\rm MeV}$ and for CMB is somewhere around $T\sim 0.1\,{\rm eV}$. If we call the energy density of the additional radiation component $\rho_{\rm dr}$, where “dr” stands for dark radiation, the presents bounds are roughly
\begin{align}\label{eq:BBNGWs}
\frac{\rho_{\rm dr}}{\rho_{\rm rad}}\Big|_{\rm CMB/BBN}\sim 0.1\,,
\end{align}
here $\rho_{\rm rad}$ is only due to the relativistic particles of the standard model. In the future, experiments such as CMB-S4 might improve \eqref{eq:BBNGWs} down to $0.01$. This will be a great opportunity to test the presence of additional relativistic particles.\footnote{
Very often, not to say always, the effects of extra are recast as having additional number of relativistic species. For historical reasons, one introduces $\Delta N_\nu$ new neutral Fermion particles ($\nu$ because the initial interest was in neutrinos) and computes the maximum allowed $\Delta N_\nu$ by BBN and/or CMB. If you check the paper \cite{Aghanim:2018eyx}, the constraints are given in terms of a quantity called $N_{\rm eff}$, the effective number of relativistic species at the relevant times. } 

Let us place an upper bound on how much GWs there can  be. Since in Eq.~\eqref{eq:GWstoday} we wrote $\Omega_{\rm GW}$ in terms of $\Omega_{\rm rad}$, we can take $10\%$ of that which means
\begin{align}\label{eq:BBNbounds}
\Omega^{\rm tot}_{\rm GW,0}h^2<10^{-6}.
\end{align}
Note that this is an integrated effect as it probes the total energy density in GWs at that time. This means we have to compute $\Omega^{\rm tot}_{\rm GW,0}=\int_{f_{\rm cut}}^\infty df \,\Omega_{\rm GW,0}(f)$, where $f_{\rm cut}\sim f_{\cal H}$ is the frequency \eqref{eq:f0s} corresponding to the Hubble radius at the time of interest. For BBN we have $f_{\cal H}\sim 10^{-12}\,\rm Hz$ and for CMB $f_{\cal H}\sim 10^{-17}\,\rm Hz$. More conservatively, tensor modes are not GWs right at Hubble radius crossing, so  one should take a cut-off at $f_{\rm cut}\sim 0.01 f_{\cal H}$ \cite{Smith:2006nka}.
\end{tcolorbox}
\newpage

Note that we did not consider yet that some PBHs evaporated long ago. Let us quickly compute the boundaries between tiny (= evaporated) and heavy (= still present today) PBHs. After formation, PBHs lose mass by Hawking evaporation. The mass loss rate is given by
\begin{align}\label{eq:hawkingevaporation}
\frac{dM_{\rm PBH}}{dt}=-\frac{A M_{\rm pl}^4}{M_{\rm PBH}^2}
\qquad {\rm where}\qquad
A=\frac{3.8\pi g_H(T_{\rm PBH})}{480}\,,
\end{align}
where $t$ is the cosmic time and $g_H$ are the spin-weighted degrees of freedom (for the PBH it is easier to create a particle with less spin so $g_{H,i}$ for each component will be larger the smaller the spin). The temperature associated to the PBH reads
\begin{align}
T_{\rm PBH}\equiv M_{\rm pl}^2/M_{{\rm PBH}}\approx 10^9{\rm GeV}\left(\frac{M_{\rm PBH}}{10^4{\rm g}}\right)^{-1}\,.
\end{align}
With the evaporation rate \eqref{eq:hawkingevaporation} we can show that the time of evaporation is given by
\begin{align}\label{eq:teva}
t_{\rm eva}\approx \frac{160}{3.8\pi g_H(T_{\rm PBH})}\frac{M_{\rm PBH,f}^3}{M_{\rm pl}^4}\approx 400\,{\rm qs} \left(\frac{M_{\rm PBH,f}}{1\,{\rm g}}\right)^{3}\,,
\end{align}
where ${\rm qs}=10^{-30}{\rm s}$ and it is called a quectosecond. I also used $g_H(T_{\rm PBH})\approx 108$ which is true for very light PBHs and the standard model of particle physics. For future reference, we have that
\begin{align}
M_{\rm PBH}(t)\approx M_{\rm PBH,f}\left(1-\frac{t}{t_{\rm eva}}\right)^{1/3}\,,
\end{align}
and that
\begin{align}
\rho_{\rm PBH}\approx M_{\rm PBH}(t)n_{\rm PBH}(t)=\rho_{\rm PBH,f}\left(\frac{a}{a_f}\right)^{-3}\left(1-\frac{t}{t_{\rm eva}}\right)^{1/3}\,.
\end{align}
In the equation above I used that since PBHs are not created or destroyed (not until evaporation at least) the number density of PBHs is conserved, namely that $n_{\rm PBH}\propto a^{-3}$ (it dilutes as the volume).

Let me discuss some relevant mass ranges. First, perhaps you are familiar with the standard estimate that says that PBHs with $M_{\rm PBH,f}>10^{15}{\rm g}$ have a lifetime longer than the age of the Universe. This is what you get by equating \eqref{eq:teva} to the age of the universe ($\approx 14 \rm Gyr$). Interestingly, asteroid mass PBHs ($M_{\rm PBH,f}\sim 10^{-15}-10^{-12}M_\odot$) can make up for the totality of dark matter. PBH with masses larger than $10^{4}M_\odot$ need to be quite or very subdominant not to get into trouble with observations. A similar thing occurs for $10^9{\rm g}<M_{\rm PBH,f}<10^{15}{\rm g}$: although they already evaporated they can inject too much energy at times where we can check it, such as BBN and CMB. Lastly, one very interesting mass range is $M_{\rm PBH,f}<10^{9}{\rm g}$ which quite remarkably we can only test with GWs.\\

In what follows, we will briefly go through each GW signal by order of my arbitrary relevance:
\begin{itemize}
\item[\S~\ref{sub:mergers}]  GWs from PBH mergers,
\item[\S~\ref{sec:igws1}] GWs induced by primordial fluctuations,
\item[\S~\ref{sub:densityfluct}] GWs induced by PBH number density fluctuations,
\item[\S~\ref{sub:Hawking}]  GWs from Hawking evaporation.
\end{itemize}

\newpage

\subsection{GWs from PBH mergers \label{sub:mergers}}

Although this topic actually deserves a whole set of lectures, I will try to explain the main features. To those interested in learning more, I suggest to check \cite{Domenech:2021odz} for a short paper with a collection of the relevant references.

PBHs are formed in the radiation dominated early universe. PBH formation is a rare event so in the beginning there is barely one PBH per Hubble patch. As the universe expands, there is eventually a period when there is more than one PBH per Hubble patch. To convince yourselves, take the inter-PBH separation \eqref{eq:interfhpbh} but evaluated at an arbitrary redshift $z$ and multiplied by the Hubble radius at that time, namely
\begin{align}
\frac{1/H}{d_{\rm PBH}}=\frac{a_{\rm f}}{{\cal H}_{\rm f}d_{\rm PBH,f}}\frac{a}{a_{\rm f}}=\frac{f_{\rm inter-BH,0}}{f_{{\cal H}_{\rm f},0}}\frac{a}{a_{\rm f}}\sim \beta^{1/3}\frac{a}{a_{\rm f}}\,,
\end{align}
where we used $d_{\rm PBH}\sim n_{\rm PBH}^{-1/3}$. This measure how many PBHs we have per Hubble length. We see that after $a/a_{\rm f}\sim \beta^{-1/3}$ there will typically more than one PBH per Hubble patch. We can also write it in terms of ${\rm f}_{\rm PBH}=\rho_{\rm PBH}/\rho_{\rm CDM}$. This gives
\begin{align}
\frac{1/H}{d_{\rm PBH}}\approx \frac{3\times 10^9}{1+z}{\rm f}^{1/3}_{\rm PBH}\left(\frac{M_{\rm PBH,f}}{M_\odot}\right)^{-1/3}\left(\frac{g_\rho(T_{\rm f})}{10.75}\right)^{-1/4}\left(\frac{g_s(T_{\rm f})}{10.75}\right)^{1/3}\,.
\end{align}
This means that even for solar mass PBHs there is a time in radiation domination where there are more than one PBH per Hubble patch. It turns out that if ${\rm f}_{\rm PBH}$ is not too small (${\rm f}_{\rm PBH}>10^{-15}$) most of the binaries form during radiation domination. \\

\begin{tcolorbox}[title=\bf{Exercise},colback=white,colframe=black!50]
Using the numbers in Tab.~\ref{tab:quantities}, the definition of $\beta$ and ${\rm f}_{\rm PBH}$ and \eqref{eq:zf}, show that
\begin{align}
\beta&={\rm f}_{\rm PBH}\,\Omega_{\rm CDM,0}h^2\,\left(\frac{H_{100}}{H_{\rm f}}\right)^2\left(\frac{a_{0}}{a_{\rm f}}\right)^3\nonumber\\&
\approx 3\times 10^{-9}\,{\rm f}_{\rm PBH}\left(\frac{M_{\rm PBH,f}}{M_\odot}\right)^{1/2}\left(\frac{g_\rho(T_{\rm f})}{10.75}\right)^{-3/4}\left(\frac{g_s(T_{\rm f})}{10.75}\right)\,.
\end{align} 
See how for ${\rm f}_{\rm PBH}<1$ the initial fraction of PBHs $\beta$ is quite small.
\end{tcolorbox}

The way in which these PBH binaries form is as follows. As the Universe expands, the closest PBHs will feel the gravitational pull towards each other. As they barely have any initial velocity, they will start course to head-on collision. But, the third closest PBHs is there to save the day and place some torque on the system. This leads to very eccentric binaries. Under these assumptions, one can find that the merger rate per unit of time and volume is given by \cite{Sasaki:2018dmp} 
\begin{align}\label{eq:mergerrate}
{\cal R}_{\rm merger}=\frac{dN_{\rm merger}}{dtdV}\approx 4\times 10^6\,\text{Gpc}^{-3}\,\text{yr}^{-1}\frac{
   {\rm f}_{\rm PBH}^2}{\left({\rm f}_{\rm PBH}^2+\sigma_{\rm eq}^2\right)^
   {21/74} 
   }\left(\frac{M_{\rm PBH,f}}{M_\odot}\right)^{-32/37}\,,
\end{align}
where $\sigma_{\rm eq}^2\approx 2.5\times 10^{-5}$.\footnote{Eq.~\eqref{eq:mergerrate} is valid as long as ${\rm f}_{\rm PBH}\ll1$, maybe ${\rm f}_{\rm PBH}<10^{-2}$. If the fraction of PBHs is high, N-body interactions become important.} Here I assumed that $M_{\rm PBH,f}>10^{15}\,{\rm g}$ so that PBHs survive until today. These mergers occur in the nearby Universe and, therefore, we can estimate the maximum frequency as that of the Innermost Stable Circular Orbit (ISCO) at a distance of $r_{\rm ISCO}=6G(M_{{\rm PBH},1}+M_{{\rm PBH},2})$, namely
\begin{align}
f^{\rm max}_{\rm GW,binary}\sim 2f_{\rm ISCO}= \frac{1}{2\pi}\frac{1}{\sqrt{6}r_{\rm ISCO}}\sim 4.4\,{\rm kHz}\left(\frac{M_{\rm PBH,f}}{M_\odot}\right)^{-1}\,,
\end{align}
where the $\sqrt{6}$ comes from using Kepler’s law, i.e. orbital frequency of a circular orbit is $\omega^2=GM_{\rm PBH,f}/r^3$.
Note that $f^{\rm max}_{\rm GW,binary}=\tfrac{2}{3\sqrt{6}}f_{r_{\rm PBH}}$ so that our estimate \eqref{eq:frpbh} was not too far off. Also note that we calculated the frequency in the source frame, so for the measured frequency we need to include the redshift as in \eqref{eq:frpbh}. Unfortunately for PBH dark matter, binaries of PBHs lighter that a solar mass lead to GWs with too high frequency to be detected. PBHs with tenths of solar masses may explain part of the LIGO/VIRGO events. To explain the LIGO/VIRGO events, which has a detected rate of around $10\,\,\text{Gpc}^{-3}\,\text{yr}^{-1}$ we need ${\rm f}_{\rm PBH}\sim 10^{-3}$.

\subsubsection{GW background from unresolved binaries}

There is also the possibility that binaries are too far away to be resolved. Note that this also includes PBHs that evaporated. In that case, the GWs from unresolved binaries form a GW background. To compute this, we need to know the GW energy radiated per binary per frequency and integrate it over the whole history knowing how often they merge. In formulas, this reads
\begin{align}
\Omega_{\rm GW,binary}=\frac{1}{3H_0^2M_{\rm pl}^2}\int_0^{f_{\rm max}/f-1}dz\frac{{\cal R}_{\rm merger}}{H(z)}f_s\frac{dE_s}{df_s}\,,
\end{align}
where $f_s$ is the source frequency, so $f_s=f/a$, and the $H(z)$ appears because $dt=dz/H$. The upper cut-off of the integral at $z_{\rm max}=f_{\rm max}/f-1$ is due to the fact that after the binary merges, there are no more GWs to be detected. You can find all the formulas and in particular that of ${dE_s}/{df_s}$ in \cite{Wang:2019kaf} (and references therein). To get a grasp of the amplitude, you can use the mean value theorem to evaluate the integral as the integrand at some intermediate value of $z$. Knowing that the closest binaries will contribute highest close the peak frequency, you can look for the maximum of integrand by neglecting the redshift and setting $f_s=\alpha f_{\rm max}$. Then, find the value of $\alpha$ which maximizes the integrand. By doing so, I got
\begin{align}
\Omega^{\rm max}_{\rm GW,binary}h^2\approx 10^{-8}\left(\frac{M_{\rm PBH,f}}{M_\odot}\right)^{5/37}\left(\frac{{\rm f}_{\rm PBH}}{0.01}\right)^{53/37}\,.
\end{align}
After the peak the GW spectrum falls off at low frequencies as $f^{2/3}$. To compute the GW background from PBHs that once dominated the universe see the appendix of \cite{Inomata:2020lmk}. These GWs are however too high frequency and I will not consider them further.

\newpage

\subsection{GWs induced by primordial fluctuations \label{sec:igws1}}

Whenever we have primordial fluctuations (scalar, vector or tensor in general), there will inevitably be a secondary generation of GWs. Mathematically speaking, this must be a secondary effect because of the decomposition theorem. But, intuitively, such inhomogeneities lead to non-vanishing anisotropic stress (in the expansion \eqref{eq:expansion2} this partly comes from considering $G^{(2)}_{\mu\nu}[\eta+h]$ as a source to linear GWs). I will refer to this effect as \textit{induced} GWs (the first works on induced GWs are \cite{Tomita,Matarrese:1992rp,Matarrese:1993zf}). Then, the question is not so much whether GWs are generated but whether GWs are generated with high enough amplitude. This includes two possibilities: $(i)$ primordial fluctuations are large and $(ii)$ the expansion history is such that the production of induced GWs is enhanced.

In \S~\ref{sec:4}, I will focus on the case of scalar induced GWs after inflation during radiation domination. We will go through some estimates there. Here, I would like to list other interesting possibilities:
\begin{itemize}
 \item GWs during axion inflation induced by gauge fields. This induces GWs during inflation which are parity violating, i.e. there is more of one helicity (left or right) than the other. I recommend you have a look at this review paper in Nature by Eiichiro Komatsu \cite{Komatsu:2022nvu}.
 \item GWs induced during inflation by scalar fields. This requires several fields contributing to boost the amplitude or resonances.
 \item GWs induced during an early matter phase with a sudden transition \cite{Inomata:2019ivs}. We will look into this schematically in the next subsection \ref{sub:densityfluct}. 
\end{itemize}

\subsection{GWs induced by PBH number density fluctuations \label{sub:densityfluct}}

This case is a bit unexpected, at least it was to me. I have argued in \S~\ref{sec:igws1} that whenever we have density fluctuations (I called them scalar fluctuations then), there will be generation of induced GWs. Which density fluctuations are the source of induced GWs in this case? For this we have to understand/recall a bit better PBH formation. 

PBH formation is a rare event in the Universe. Only a tiny fraction of Hubble patches will have a fluctuation large enough to collapse and form a PBH. We can also say that to a good approximation PBH formation will occur randomly in space according to a uniform distribution (i.e. each patch in space has the same probability to have a PBH). This means that \textit{on average} PBH number density is homogeneous. But, it is a random event and there will be fluctuations. In particular, there will be inhomogeneities in the number density of PBHs across space. When we have a rare, discrete, uniform distribution, the fluctuations are what is called “white noise” and follow a Poisson distribution (i.e. the intensity of the noise or the root mean square is independent of frequency).\footnote{For example, in a homogeneous city, with all crossroads equally dangerous (or safe), the spatial distribution of car accidents would follow the Poisson statistics, unless the drivers are actively motivated to crash into each other. }

We will then treat the gas of PBHs as a perfect fluid which is homogeneous on average but has some density fluctuations. Now, because the distribution of PBHs is discrete, fluctuations have a cut-off at the mean inter-PBH separation, what we called $d_{\rm PBH,f}$ in \eqref{eq:interfhpbh}. This cut-off is just to signify that below the mean inter-PBH separation things are totally discrete: either you see a PBH or you do not (and you probably do not). As a Poisson spectrum, the density fluctuations spectrum  $\langle \delta n_{\rm PBH}/n_{\rm PBH}\rangle\sim \rm constant$, and the dimensionless spectrum (see below Eq.~\eqref{eq:hijhijdimension}) is given by
\begin{align}
{\cal P}_{\delta n_{\rm PBH}/n_{\rm PBH}}\sim {\cal O}(0.1)\left(\frac{k}{k_{\rm UV}}\right)^3\,
\end{align}
where $k_{\rm UV}=a_f/d_f$ is the high momentum (or UV) cut-off. What is important is that the dimensionless spectrum peaks at $k_{\rm UV}$ where it is also cut off. These PBH density fluctuations are then a source of induced GWs. One of the caveats is that unless PBHs dominate the early universe at some point, the resulting induced GWs are suppressed by the ratio $\rho_{\rm PBH}/\rho_{\rm rad}$. So in what follows I will assume that PBHs dominate the universe.\\

\begin{tcolorbox}[title=\bf{Exercise},colback=white,colframe=black!50]
Consider that you form PBHs with mass $M_{\rm PBH,f}$ with initial fraction $\beta$ in a radiation dominated universe. Show that for
\begin{align}
\beta>\beta_{\rm min}\approx  6\times 10^{-6}\left(\frac{M_{\rm PBH,f}}{1\,{\rm g}}\right)^{-1}\,,
\end{align}
PBHs dominate the very early universe. To do this you have to estimate at which time PBHs have the same energy density as radiation (some PBH-radiation equality) and require that such equality time occurs before the time of evaporation \eqref{eq:teva}.

\end{tcolorbox}

If PBHs dominate the universe, they eventually reheat it by Hawking evaporation. We know at what time they evaporate, Eq.~\eqref{eq:teva}. Assuming a monochromatic mass function, all PBHs evaporate at the same time. The final moments between PBH domination and radiation domination is almost instantaneous: the final glow of the PBHs is what eventually reheats the universe. In practice though, it takes around $1/4$ of an e-folding (i.e. $\ln (a_{\rm eva}/a)\sim 1/4$). Since we know the time of evaporation, we also know the expansion rate at that time by the Friedmann equations, $H_{\rm eva}=2/3/t_{\rm eva}$. Using the formulas in App.~\ref{app:formulasuseful}, we can compute the temperature of the radiation filling the universe at that time, namely
\begin{align}
T_{\rm eva}\approx
2.8\times 10^{10}\,{\rm GeV}\,\left(\frac{M_{\rm PBH,f}}{1\,{\rm g}}\right)^{-3/2}\left(\frac{g_{\rho}(T_{\rm eva})}{106.75}\right)^{-1/4}\,.
\end{align}
Once we know the temperature at evaporation, we can compute the redshift of evaporation similarly as we did in \eqref{eq:zf}. This time we find
\begin{align}
1+z_{\rm eva}\approx 10^{23}\left(\frac{M_{\rm PBH,f}}{1\,{\rm g}}\right)^{-3/2}\left(\frac{g_{s}(T_{\rm eva})}{106.75}\right)^{1/3}\left(\frac{g_{\rho}(T_{\rm eva})}{106.75}\right)^{-1/4}\,.
\end{align}

Regarding the GW signature, I will only present the final results and refer the interested reader to Refs.~\cite{Domenech:2020ssp} and references therein. You can also check \cite{Domenech:2023mqk} for a lighter version (we also considered that PBHs leave remnants and become the dark matter). The cut-off frequency today \eqref{eq:interfhpbh} taking into account that PBHs dominated the universe now reads
\begin{align}
f_{\rm UV,0}
\approx 3.6\times 10^{6}\, {\rm Hz}\,\left(\frac{M_{\rm PBH,f}}{1\,{\rm g}}\right)^{-5/6}\left(\frac{g_{\rho}(T_{\rm eva})}{106.75}\right)^{1/4}\left(\frac{g_{s}(T_{\rm eva})}{106.75}\right)^{-1/3}\,.
\end{align}
The GW spectral density is then given by
\begin{align}\label{eq:Omegaeva}
  \Omega_{\rm GW,eva}\approx \left(\frac{k}{k_{\rm UV}}\right)^{11/3}\Omega^{\rm peak}_{\rm GW,eva}\Theta(k_{\rm UV}-k)\,,
\end{align}
where $\Theta(k_{\rm UV}-k)$ is the Heaviside theta and
\begin{align}\label{eq:Omegapeak}
  \Omega^{\rm peak}_{\rm GW,eva} \approx\frac{1}{10} \left(\frac{\beta}{10^{-3}}\right)^{16/3}\left(\frac{M_{\rm PBH,f}}{1\,{\rm g}}\right)^{34/9}\,.
\end{align}
The reason why the GW spectrum gets such a high amplitude is because during PBH domination (= pressureless matter domination) density fluctuations grow but do not propagate. Actually, they grow quite a lot to the point that some of them enter the non-linear regime. However, the almost sudden PBH evaporation transform such big density fluctuations into big radiation fluctuations. And radiation moves fast. This causes huge velocity wakes in the radiation fluid which is a big source for induced GWs. The most interesting part is that, within some caveats, we can constrain the maximum amount of PBHs at formation. Requiring that \eqref{eq:Omegapeak} satisfies current BBN bounds (see Eq.~\eqref{eq:BBNbounds}) we obtain
\begin{align}\label{eq:betamax}
\beta < 10^{-3}\left(\frac{M_{\rm PBH,f}}{1\,{\rm g}}\right)^{-17/24}\,.
\end{align}
Although these estimates are  sensitive to the monochromatic mass assumption and, perhaps, the dynamics of the non-linear regime, they constitute promising and perhaps the only way to test PBH dominated epochs.

\subsection{GWs from Hawking evaporation \label{sub:Hawking}}

Our last GW signature from PBHs are gravitons from Hawking evaporation itself. This is quite an interesting possibility. As you may expect, this GW signature is likely only important if PBHs dominate the universe such that PBHs products dominate as much as possible after evaporation. Let us first estimate the peak frequency of the graviton spectrum. In fact, we have almost computed it in \eqref{eq:frpbh}. Very roughly, what happens is that particles are emitted with a typical physical momentum proportional to the temperature of the BH at the time of emission, namely $k_{\rm phys}\sim a(t)T_{\rm PBH}(t)$. But, the mean energy density of PBHs decays as $a^{-3}$ while the energy density of relativistic particles decays as $a^{-4}$. This means that particles emitted early on do not contribute much to the total energy density of relativistic particles at evaporation. On the other hand, very close to evaporation we have that $T_{\rm PBH}(t)$ grows very quickly but $\rho_{\rm PBH}$ also decays very quickly. This means that, as we expect, extremely high energy particles emitted at the last instants of PBH evaporation when the PBH mass is almost Planckian, barely contribute to the total energy density of radiation after evaporation. We can conclude that most of the contribution to the energy density will come from particles emitted close to evaporation but not too close. Since the final evaporation happens very quickly in terms of cosmic times, it is a good approximation to neglect the dependence of $t$ in the temperature and evaluate the physical momentum at evaporation, namely $k_{\rm typical}\sim a_{\rm eva}T_{\rm PBH,f}$. Thus, the typical frequency of the gravitons will be proportional to the initial PBH temperature. Putting some numbers and using that PBHs dominated the universe we have
\begin{align}\label{eq:fgraviton}
f_{\rm graviton,0}=f_{r_{\rm PBH},0}\approx 10^{14}\,{\rm Hz}\left(\frac{M_{\rm PBH,f}}{1\,{\rm g}}\right)^{1/2}\left(\frac{g_{\rho}(T_{\rm eva})}{106.75}\right)^{1/4}\left(\frac{g_{s}(T_{\rm eva})}{106.75}\right)^{-1/3}\,.
\end{align}
For more details on the calculations of the spectrum I recommend you look at appendix A of Ref.~\cite{Inomata:2020lmk}. Unfortunately, the peak frequency of the graviton spectrum \eqref{eq:fgraviton} is a very high frequency signal for the GW background. It can only be probed by their contribution to the radiation energy density in future experiments like CMB-S4. 

To compute the graviton spectrum we need to take into account the energy emission rate of such particles by Hawking evaporation and integrate it throughout the whole PBH lifetime. The appendix of Ref.~\cite{Inomata:2020lmk} explains this quite well. We can, however, use some hack. All we need to know is that each particle is emitted with the same momentum distribution (i.e. a thermal distribution) but with different coefficients depending on the spin of the particle (the spin-weighted degrees of freedom $g_{s_i,H}$ for an arbitrary spin $s_i$). Below Eq.~\eqref{eq:teva} we said that the total spin-weighted degrees of freedom $g_{H}(T_{\rm PBH})\approx 108$ for tiny PBHs. The spin-weighted degrees for a spin 2 particle (like the graviton) for a non-spinning black hole (if it spins it is more likely to emit a particle with spin) is $g_{s=2,H}\approx 0.1$. This means that right after evaporation we have
\begin{align}
\frac{\rho_{\rm graviton}}{\rho_{\rm rad}}\Big|_{\rm eva}= \frac{g_{s=2,H}}{g_{H}(T_{\rm PBH})}\approx 10^{-3}\,.
\end{align}
Note that if we want to compare it to CMB or BBN bounds we should follow the energy ratio for the gravitons $\Omega_{\rm GW,graviton}$ until the relevant times. This would change a bit the factor $10^{-3}$ we derived.  Sadly, even doing so, this is a bit below what an experiment like CMB-S4 could reach. But, if we have highly spinning PBHs the fraction could get to ${\cal O}(0.1)$ which is definitely within range. Assuming the standard model of particle physics, all the other products of PBH evaporation will be non-relativistic by the time of BBN, except for photons, neutrinos and gravitons. It is exciting that we can test PBH dominated universe with the GWs from Hawking evaporation.\\

\begin{tcolorbox}[title=\bf{Try it!},coltitle=black,colback=white,colframe=orange!50]
Download the latest version of \href{https://blackhawk.hepforge.org}{BlackHawk} by Alexandre Arbey and Jérémy Auffinger and compute the resulting spectrum of gravitons from Hawking evaporation in the blink of an eye. Check also the arXiv paper \href{https://arxiv.org/abs/2108.02737}{here}.
\end{tcolorbox}

\newpage 

\section{The case of scalar induced GWs \label{sec:4}}

To end the lectures, I would like to enter into a bit of detail with the scalar induced GWs. I will not derive the general formulas as this can be found in, e.g., my review paper \cite{Domenech:2021ztg}. I will focus on understanding the basic steps in the calculations and the main physics of induced GWs. 

Let us start with the formal equation for induced GWs. If we go to second order in perturbation theory we can write that the linear tensor modes now have a source term, namely
\begin{align}
h_{ij}''+2{\cal H}h_{ij}'-\Delta h_{ij}=\widehat{TT}^{kl}_{ij}\left[T^{(2)}_{kl}-G^{(2)}_{kl}[\eta+h]\right]\,,
\end{align}
where $\widehat{TT}^{kl}_{ij}$ is the transverse-traceless projector defined in App.~\ref{app:decomposition}. If we only focus on purely scalar perturbations, the source term must be bilinear in scalar. We can simplify the calculations by neglecting the purely geometrical term $G^{(2)}_{kl}[\eta+h]$ and focus only on $T^{(2)}_{kl}$. This is more or less justified since on sub-Hubble scales matter fluctuations dominate over metric fluctuations. As we shall see later though, induced GWs are mostly generated on scales close to the Hubble radius, so neglecting the geometrical term induces some ${\cal O}(1)$ error but not more.\\

\begin{tcolorbox}[title=\bf{Try it!},coltitle=black,colback=white,colframe=orange!50]
Use \textsc{Mathematica} package \href{http://www2.iap.fr/users/pitrou/xpand.htm}{\textsc{xPand}} to check the second order equations for $h_{ij}$. You can pick any gauge fixing you like, but the Newton gauge (where $E=B=0$ in Eq.~\eqref{eq:perturbFLRW}) is the simplest. You should get
\begin{align}
h_{ij}''+2{\cal H}h_{ij}'-\Delta h_{ij}=\widehat{TT}^{kl}_{ij}\left[4\partial_k\phi\partial_l\phi+2(\rho+p)\partial_k v\partial_l v\right]\,,
\end{align}
where $v$ is the perturbation of the spatial velocity of the fluid (see the discussion below Eq.~\eqref{eq:tmunu1}). You may also have to use the first order equations for the scalar components, specially that $\phi+\psi=0$ in the absence of shear. Also note that for some reason \textsc{xPand} defines $H_{ij}\supset 2 h_{ij}$. So compared to Eq.~\eqref{eq:perturbFLRW} there is an additional factor $2$ you have to take care of to recover the results in this lecture.
\end{tcolorbox}

To make things even more simple, we consider the fluctuations of a massless scalar field, say $\delta\varphi$. From its energy momentum tensor given in \eqref{eq:tmunu2}, the equations of motion for induced GWs read
\begin{align}\label{eq:induced1}
h_{ij}''+2{\cal H}h_{ij}'-\Delta h_{ij}=\widehat{TT}^{kl}_{ij}\left[\partial_k\delta\varphi\partial_l\delta\varphi\right]\,,
\end{align}
which is as simple as it gets. Let us work with Fourier modes. For the tensors the Fourier expansion is given in Eq.~\eqref{eq:hijfourier}. For the scalar fluctuations we have
\begin{align}\label{eq:fourierphi}
\delta\varphi=\frac{1}{(2\pi)^3}\int d^3q \,\delta\varphi_{\bm{q}}(\tau)\, e^{i\bm{q}\cdot\bm{x}}\,.
\end{align}
Reality of $\delta\varphi$ implies that $\delta\varphi^*_{\bm{q}}=\delta\varphi_{-\bm{q}}(\tau)$. This won't matter much for us though. With these expansions, Eq.~\eqref{eq:induced1} becomes
\begin{align}
\frac{1}{(2\pi)^3}\sum_{\tilde\lambda}&\int d^3\tilde k \,\epsilon_{ij}^{\tilde\lambda}(\tilde{\bm{k}})\, e^{i\tilde{\bm{k}}\cdot\bm{x}}\left\{h''_{\tilde{\bm{k}},{\tilde\lambda}}+2{\cal H}h'_{\tilde{\bm{k}},{\tilde\lambda}}+\tilde k^2h_{\tilde{\bm{k}},{\tilde\lambda}}\right\}\nonumber\\&=-\frac{1}{(2\pi)^6}\int d^3q_1 d^3q_2 \,\widehat{TT}^{kl}_{ij}[\bm{q}_1+\bm{q}_2](\bm{q}_1)_k(\bm{q}_2)_l\,\delta\varphi_{\bm{q}_1}\delta\varphi_{\bm{q}_2}\, e^{i(\bm{q}_1+\bm{q}_2)\cdot\bm{x}}\,.
\end{align}
We are still half way, so let us do the Fourier inverse by multiplying by $\int d^3x \,(\epsilon^{ij}_\lambda({\bm{k}}))^* e^{-i{\bm{k}}\cdot\bm{x}}$ and integrate over $\bm{x}$. The integral over $\bm{x}$ is rather easy at it gives $(2\pi)^3\delta(\tilde{\bm{k}}-{\bm{k}})$ and $(2\pi)^3\delta(\bm{q}_1+\bm{q}_2-{\bm{k}})$, so that after integration $\tilde{\bm{k}}={\bm{k}}=\bm{q}_1+\bm{q}_2$. Then, we can also use that  $\epsilon^{ij}_\lambda({\bm{k}})\widehat{TT}^{kl}_{ij}[\bm{q}_1+\bm{q}_2]=\epsilon^{ij}_\lambda({\bm{k}})$ after imposing the Dirac deltas. You may explicitly check this if you want but since $\widehat{TT}^{kl}_{ij}$ is the transverse-traceless projector, the transverse-traceless projection of $\epsilon^{ij}_\lambda({\bm{k}})$ is $\epsilon^{ij}_\lambda({\bm{k}})$. We are left to use the normalization and reality conditions $\epsilon^{ij}_\lambda({\bm{k}})$ and that is it. All in all, we arrive at
\begin{align}
h''_{{\bm{k}},{\lambda}}+2{\cal H}h'_{{\bm{k}},{\lambda}}+ k^2h_{{\bm{k}},{\lambda}}=-\frac{1}{(2\pi)^3}\int d^3q_1  \,\epsilon^{kl}_\lambda(-{\bm{k}})(\bm{q}_1)_k({\bm{k}}-\bm{q}_1)_l\,\delta\varphi_{\bm{q}_1}\delta\varphi_{{\bm{k}}-\bm{q}_1}\,.
\end{align}
I will do one last cosmetic touch by defining $\bm{q}_1=\bm{q}$ and using that by orthogonality $\epsilon^{kl}_\lambda({\bm{k}})({\bm{k}}-\bm{q})_l=-\epsilon^{kl}_\lambda({\bm{k}})(\bm{q})_l$. Finally, the equations of motion for scalar induced GWs in Fourier space are given by
\begin{align}\label{eq:inducedfinal}
h''_{{\bm{k}},{\lambda}}+2{\cal H}h'_{{\bm{k}},{\lambda}}+ k^2h_{{\bm{k}},{\lambda}}=\frac{1}{(2\pi)^3}\int d^3q  \,\epsilon^{kl}_\lambda(-{\bm{k}})q_kq_l\,\delta\varphi_{\bm{q}}\delta\varphi_{{\bm{k}}-\bm{q}}\equiv {\cal S}_{\bm{k},\lambda}(\tau)\,.
\end{align}
You can compute the product of the polarization tensors with the formulas in App.~\ref{app:polarization} or check \cite{Domenech:2021ztg}.

The standard way of proceeding would be to split the scalar fluctuations into a primordial (random) value and a transfer function $T_{\delta\varphi}$, something like
\begin{align}
\delta\varphi_{\bm{q}}=\delta\varphi^{\rm prim}_{\bm{q}}\times T_{\delta\varphi}(q\tau)\,.
\end{align}
The transfer function is found by solving the scalar field linear equations of motion. We can then use the Green’s method to find the formal solution to \eqref{eq:inducedfinal}. Just in case, the Green’s method tells us that if we know the homogeneous solutions to the equation (which we know, see Eq.~\eqref{eq:generalh}), we can find the particular solution by integrating the source term times the Green’s function. Once we find the formal solution we can compute the two point function for the tensor modes and the predicted spectral density of induced GWs.

Let us do what I said above formally and schematically. Let me call $G(\tau,\tilde\tau)$ the Green’s function for the tensor modes. Then, the formal solution to Eq.~\eqref{eq:inducedfinal} is given by
\begin{align}
h_{{\bm{k}},{\lambda}}(\tau)=\int_{\tau_i}^\tau d\tilde\tau\, G(\tau,\tilde\tau)\,{\cal S}_{\bm{k},\lambda}(\tau)\,,
\end{align}
where $\tau_i$ is some initial time. 
I will not continue with the details, you can follow them, e.g., in Sec.~3.2 of \cite{Domenech:2021ztg}. What I would like you to notice is the following:
\begin{itemize}
\item The two point function of tensor modes is proportional to the four point function of scalar modes, namely
\begin{align}
\langle h_{{\bm{k}},{\lambda}}h_{{\bm{k}'},{\lambda'}} \rangle\propto \langle\delta\varphi_{\bm{q}}\delta\varphi_{{\bm{k}}-\bm{q}}\delta\varphi_{\bm{q}'}\delta\varphi_{{\bm{k}}'-\bm{q}'}\rangle\,.
\end{align}
So strictly speaking the spectrum of induced tensor modes is sourced by the trispectrum of scalar fluctuations. For the case of Gaussian fluctuations, we can write after Wick contractions
\begin{align}
\langle\delta\varphi_{\bm{q}}\delta\varphi_{{\bm{k}}-\bm{q}}\delta\varphi_{\bm{q}'}\delta\varphi_{{\bm{k}}'-\bm{q}'}\rangle=\langle\delta\varphi_{\bm{q}}\delta\varphi_{\bm{q}'}\rangle\langle\delta\varphi_{{\bm{k}}-\bm{q}}\delta\varphi_{{\bm{k}}'-\bm{q}'}\rangle+{\rm permutations}\,.
\end{align}
See Eugene Lim’s \href{https://nms.kcl.ac.uk/eugene.lim/AdvCos/lecture2.pdf}{notes} for a review of the Wick contractions. But in principle induced GWs are also sensitive to things like non-Gaussianity of primordial fluctuations (although for the effect to be significant, fluctuations have to be very non-Gaussian).

\item Since the source term \eqref{eq:inducedfinal} is only proportional to scalars, induced GWs are generated independently of whether there are primordial tensor modes or not. This means that almost always we can consider these two components separately, i.e. primordial tensor and induced tensor. I say almost always because if there are significant correlations between primordial tensor and primordial scalars there could be some mixed contribution coming from cross correlations. Do not be alarmed. There is no problem with perturbation theory. Linear tensor modes are generated only during inflation. After inflation the leading mechanism to generate tensor modes are induced GWs.

\item Induced GWs are a bit different than the typical mechanism to produce GWs. What I mean is that often GWs are generated by sub-Hubble physics (e.g. cosmic string loops, bubble collisions from phase transitions, quantum fluctuations during inflation). As we shall see, induced GWs are mostly produced at Hubble radius crossing when a given scalar modes re-enters the Hubble radius. So the GW production is truly at cosmological (Hubble radius) scales.
\end{itemize}

Before going into a toy example, we can derive some estimates for the frequency and amplitude of induced GWs. The frequency we already computed in Eq.~\eqref{eq:fhpbh}, which after putting some numbers I get
\begin{align}
f_{\rm f}=\frac{k_{\rm f}}{2\pi a_0}\approx 1.2\times 10^{8}\,{\rm Hz}\left(\frac{M_{\rm PBH,f}}{1\rm g}\right)^{-1/2}\left(\frac{g_\rho(T_{\rm f})}{106.75}\right)^{1/4}\left(\frac{ g_{s}(T_{\rm f}) }{106.75}\right)^{-1/3}\,.
\end{align}
For the amplitude of the spectral density we can be a bit sloppy and use Eqs.~\eqref{eq:spectraldensity} and \eqref{eq:GWstoday}, together with the fact that for peaked sources we expect $h\sim \delta\varphi$ so that ${\cal P}^{\rm peak}_h\sim ({\cal P}^{\rm peak}_{\delta\varphi})^2$. By doing so we arrive at
\begin{align}\label{eq:estimate}
\Omega^{\rm peak}_{\rm GW,0}\approx 10^{-5}\times \frac{1}{12}{\cal P}_h\approx 10^{-6}\left({\cal P}^{\rm peak}_{\delta\varphi}\right)^2\,.
\end{align}
We also get the same for the curvature perturbation ${\cal R}$ up to ${\cal O}(1)$ factors. Planck 2018 measured the primordial curvature perturbation on the largest scales with ${\cal P}_{\cal R}\sim 10^{-9}$. If we extrapolate this value down to the smallest scales we get $\Omega^{\rm peak}_{\rm GW,0}\sim 10^{-24}$ which is definitely not observable. But, this is just a big extrapolation. If the primordial spectrum of curvature fluctuations is enhanced on small scales, e.g. ${\cal P}_{\cal R}\sim 10^{-4}$ then we have $\Omega^{\rm peak}_{\rm GW,0}\sim 10^{-12}$ which could be seen by future GW detectors like LISA, Einstein Telescope and Cosmic Explorer. \\

\begin{tcolorbox}[title=\bf{Curiosity},colback=white,colframe=greenish!50]
Do you remember the Klein-Gordon equation? For a massless scalar field $\varphi$ it reads
\begin{align}
\tilde\nabla_\mu\tilde\nabla^\mu\varphi=\frac{1}{\sqrt{-\tilde g}}\partial_\mu\left(\sqrt{-\tilde g}\,\tilde g^{\mu\nu}\partial_\nu\varphi\right)=0\,.
\end{align}
If the scalar field $\varphi$ only has fluctuations say $\delta\varphi$, then we have that in a flat FLRW universe the equations of motion for the modes functions $\delta\phi_{\bm{q}}(\tau)$ read
\begin{align}\label{eq:deltaphi}
\delta\varphi_{\bm{q}}''+2{\cal H}\delta\varphi_{\bm{q}}'+ q^2\delta\varphi_{\bm{q}}=0\,.
\end{align}
This is the same equation that we solved in \S~\ref{sec:GWsincosmo} for the freely propagating tensor modes.
\end{tcolorbox}

\subsection{General behaviours (and important checks)}

Let us continue the discussion with a toy example. Since usually induced GWs are generated by peaked primordial spectra, let us assume that the scalar field fluctuations only depend on the modulus of the momentum and have a single Fourier mode, namely
\begin{align}
\delta\varphi_{\bm{q}}=(2\pi)\delta\varphi_{\rm peak}(\tau)\times \delta(\ln(k/k_{\rm peak}))\,,
\end{align}
where $\delta\varphi_{\rm peak}$ is some arbitrary amplitude and the logarithm inside the delta function is to keep it dimensionless. In that case, the source to induced GWs \eqref{eq:inducedfinal} reads
\begin{align}
{\cal S}_{\bm{k},\lambda}(\tau)&=\frac{1}{(2\pi)^3}\int dq \,d\cos\theta_q \, d\chi\,q^2  \,\epsilon^{kl}_\lambda({\bm{k}})q_kq_l\,\delta\varphi_{\bm{q}}\delta\varphi_{{\bm{k}}-\bm{q}}\nonumber\\&\sim \frac{\delta\varphi_{\rm peak}^2(\tau)}{2\pi}\int dq \,d\cos\theta \, d\chi\,q^4  \sin^2\theta_q \,\delta(\ln(q/k_{\rm peak}))\delta(\ln(|\bm{k}-\bm{q}|/k_{\rm peak}))\,.
\end{align}
We can relate $\cos\theta_q$ to $|\bm{k}-\bm{q}|$ by expanding the modulus, which gives
\begin{align}
\cos\theta_q=\frac{k^2+q^2-|\bm{k}-\bm{q}|^2}{2qk}\,.
\end{align}
And with that we can integrate and arrive at
\begin{align}\label{eq:ssimple}
{\cal S}_{\bm{k},\lambda}(\tau)&\sim {k_{\rm peak}^2\delta\varphi_{\rm peak}^2(\tau)}\,  \left(1-\frac{k^2}{4k^2_{\rm peak}}\right)\Theta(2k_{\rm peak}-k) \,,
\end{align}
where the Heaviside theta comes from the requirement that $\cos^2\theta\leq1$. Note that the source term vanishes at $k=2k_{\rm peak}$. This is because the scalar momenta becomes orthogonal to the polarization of the tensor modes and there is no way to generate tensor modes through scalars.
To be honest, I have cheated a little bit with the integral of the azimuthal angle $\chi_q$. The projection $\epsilon^{kl}_\lambda(-{\bm{k}})q_kq_l\sim e^{\pm 2i\chi_q}$ and would vanish after integration. That would leave the example a bit pointless. However, when computing the induced GW spectrum we will have something like $\epsilon^{kl}_\lambda(-{\bm{k}})q_kq_l\epsilon^{ij}_\lambda({-\bm{k}'})q_iq_j$ which because of momentum conservation, i.e. $\bm{k}+\bm{k}'=0$, it is independent on $\chi_q$. This is why I dropped the $\chi_q$ dependence earlier. The good thing is that now we can use \eqref{eq:ssimple} to solve for the induced tensor modes.

Now, assume radiation domination so that ${\cal H}=1/\tau$. Then, to solve for the induced tensor modes, we will use that a massless scalar field mode functions have the same solution as free tensor modes, namely Eqs.~\eqref{eq:subHphi} and \eqref{eq:superHphi}. The only thing we need to know now is that when $\delta\varphi$ is super-Hubble then it stays constant, i.e. $\delta\varphi_{\rm peak}(k_{\rm peak}\tau\ll1)\approx \delta\varphi^{\rm prim}_{\rm peak}$. When $\delta\varphi$ enters the Hubble radius it decays and becomes unimportant. For a constant source, the solution to \eqref{eq:inducedfinal} on super-Hubble scales is
\begin{align}
h_{\bm{k},\lambda}(k\tau\ll1,k_{\rm peak}\tau\ll1)\sim (k_{\rm peak}\tau)^2\,.
\end{align}
This means that while the scalar mode is super-Hubble, all super-Hubble tensor modes are growing. The growth stops when the scalar mode is sub-Hubble. After that, super-Hubble tensor modes become constant, so that if we evaluate the solution at scalar-Hubble-crossing $(k_{\rm peak}\tau=1)$ we have
\begin{align}\label{eq:superhubbletensor}
h_{\bm{k},\lambda}(k\tau\ll1,k_{\rm peak}\tau\gg1)\sim \rm constant \,.
\end{align}
After tensor modes enter the Hubble radius ($k\tau\gg1$) the behave as free tensor modes as in \eqref{eq:subHphi} with the amplitude determined at tensor-Hubble-crossing.  Note that tensor modes that enter the horizon before the scalar peak does (i.e. $k\gtrsim k_{\rm peak}$) behave slightly different. The difference is only relevant though if the sound speed of scalar field fluctuations is different than unity, e.g. $c_s^2=w$ like for an adiabatic perfect fluid. In that case, there is a sub-Hubble resonance for $k=2c_s k_{\rm peak}$ that greatly enhances the production of induced tensor modes. For $c_s^2=1$ we can get an order of magnitude estimate by evaluating \eqref{eq:inducedfinal} at horizon crossing, which yields
\begin{align}
h_{\bm{k},\lambda}(k\gtrsim k_{\rm peak})\sim \left(\frac{k_{\rm peak}}{k}\right)^2\delta\varphi_{\rm peak}^2\,.
\end{align}
Since $k<2k_{\rm peak}$, we can take that $h_{\bm{k},\lambda}(k\sim k_{\rm peak})\sim \delta\varphi_{\rm peak}^2$. This is consistent with our earlier estimate \eqref{eq:estimate}.

Let me now note that Eq.~\eqref{eq:superhubbletensor} has some interesting implications. Since the two point function of induced tensor modes \eqref{eq:hijhijdimension} is constant we find that
\begin{align}
\Omega^{\rm induced}_{\rm GW}(k)\propto {\cal P}_h^{\rm induced}(k)\propto (k/k_{\rm peak})^3\,.
\end{align}
This is the so-called universal infrared (low frequency) scaling for localized (in time) sources. In fact, for induced GWs in radiation domination it goes as $(k/k_{\rm peak})^3\ln^2(k/k_{\rm peak})$, due to some sub-Hubble residual growth, but the $k^3$ is universal. It only changes if the equation of state of the universe changes.

{\bf Important checks.}
To sum up, if you ever compute the spectrum of induced GWs you should always check that:
\begin{enumerate}
\item The low frequency tail of the spectrum decays as $(k/k_{\rm peak})^3\ln^2(k/k_{\rm peak})$, if computed in radiation domination. If not, either there is something wrong with your calculation or the primordial scalar spectrum that you are considering is not to be considered as peaked, e.g. it has a slow decay.
\item If the scalar field fluctuations propagate at a given $c_s^2\neq 1$, you should find a peak in the induced GW spectrum precisely at $k=2c_sk_{\rm peak}$. If not, it is possible that your primordial scalar spectrum should be considered as broad or $c_s=1$.
\item For log-normal like peaks in the primordial scalar spectrum, you should find a cut-off for the induced GW spectrum at $k\gtrsim 2k_{\rm peak}$. For a power-law peak in the primordial scalar spectrum, the high frequency tail should also be a power-law.
\end{enumerate}

\subsection{Let us compute: practical examples}

\begin{figure}
\centering
\includegraphics[width=0.49\columnwidth]{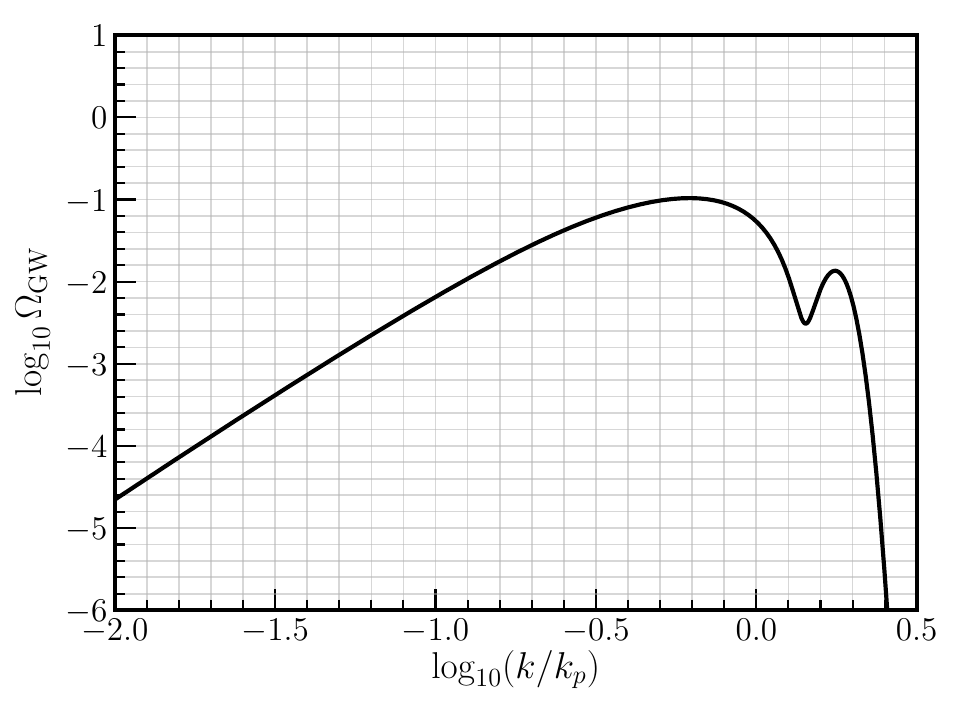}
\includegraphics[width=0.49\columnwidth]{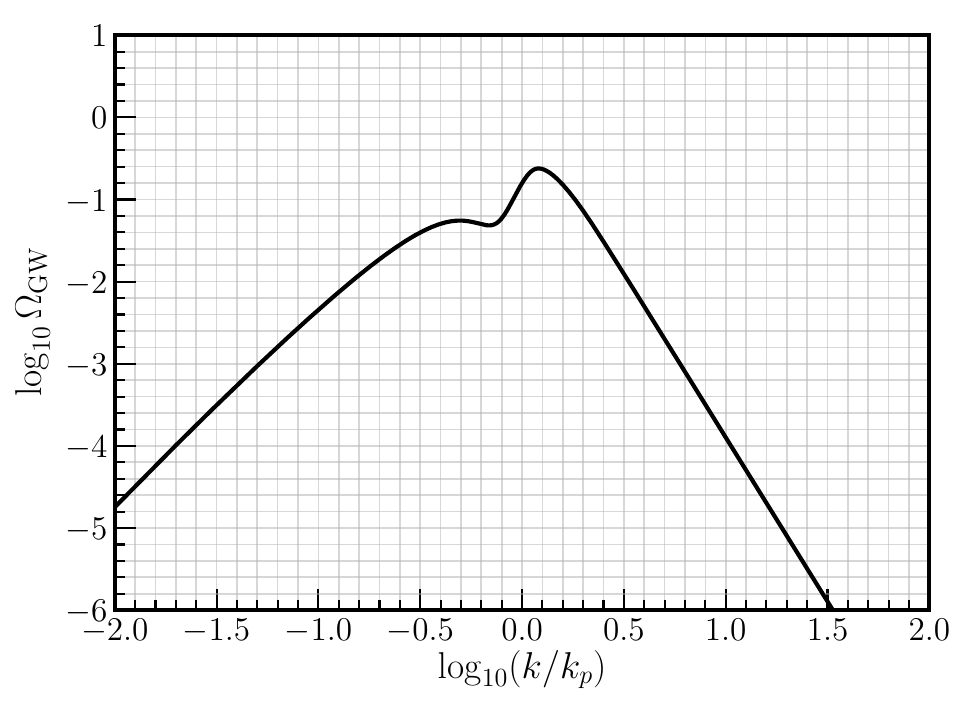}
\includegraphics[width=0.49\columnwidth]{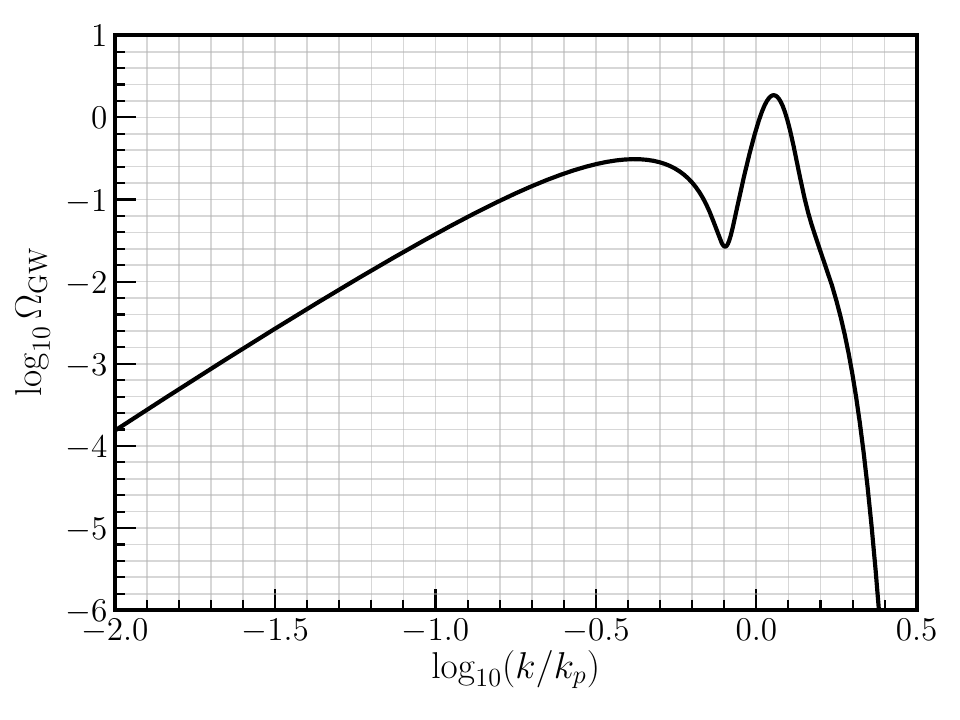}
\includegraphics[width=0.49\columnwidth]{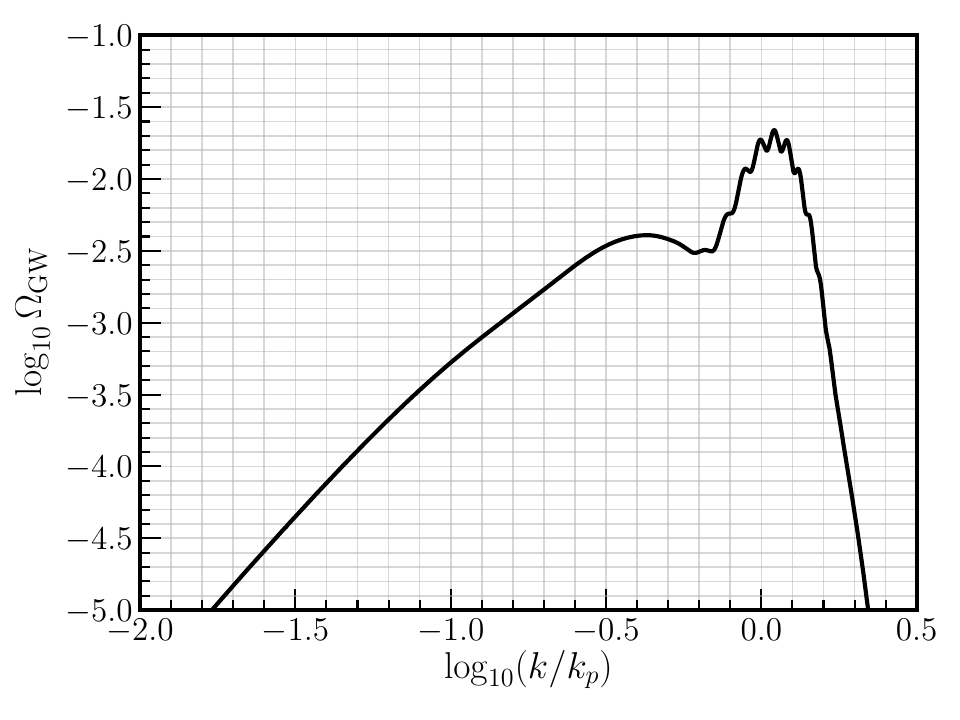}
\caption{Induced GW spectrum from various shapes of the primordial curvature power spectrum and different values of $c_s^2$. \label{fig:examples}}
\end{figure}

We now take the exact formula for induced GWs during radiation domination, which is given by
\begin{equation}\label{eq:Phgaussianfinal}
\Omega_{\rm GW,*}=\int_0^\infty dv\int_{|1-v|}^{1+v}du\,{\cal T}(u,v,c_s){{\cal P}_{\cal R}(ku)}{{\cal P}_{\cal R}(kv)}\,,
\end{equation}
where
\begin{align}\label{eq:w13}
{\cal T}_{RD}(u,v,c_s,w=1/3)=&\frac{y^2}{3c_s^4}\left(\frac{4v^2-(1-u^2+v^2)^2}{4u^2v^2}\right)^2\nonumber\\&\times
\left\{\frac{\pi^2}{4}y^2\Theta[c_s(u+v)-1]
+\left(1-\frac{1}{2}y \ln\left|\frac{1+y}{1-y}\right|\right)^2\right\}\,.
\end{align}
Remember that for an adiabatic perfect fluid we have $c_s^2=w$ and for a scalar field in an exponential potential $c_s^2=1$. In the equations above I defined $v=q/k$ and $u=|\bm{k}-\bm{q}|/k$. I computed some spectrum for you using \href{https://github.com/Lukas-T-W/SIGWfast/releases}{SIGWfast}. You can find them in Fig.~\ref{fig:examples}. Can you guess which one is which?\footnote{From left to right, top to bottom: sharp peak with $c_s^2=1$, broken power-law primordial spectrum, sharp peak with $c_s^{2}=1/3$ and sharp peak with oscillatory modulations.}

The only thing that you have left to do to become a master of the induced GWs is to be able to compute the Signal to Noise Ratio (SNR) for a given signal and a given experiment. In particular, you have to compute \cite{Schmitz:2020syl}
\begin{align}
{\rm SNR}=\left[n_{\rm det}t_{\rm obs}\int_{ f_{\rm min}}^{ f_{\rm max}}df\left(\frac{\Omega_{\rm signal}(f)}{\Omega_{\rm noise}(f)}\right)^2\right]^{1/2}\,,
\end{align}
where $n_{\rm det}=1$ or $n_{\rm det}=2$ for auto or cross-correlation of the GW signal, $t_{\rm obs}$ is the time of observation, $f_{\rm min}$ and $f_{\rm max}$ are the minimum and maximum frequencies accessible to the GW detector, $\Omega_{\rm signal}$ is the spectral density of the signal and $\Omega_{\rm noise}$ the spectral density of the noise of the detector. By requiring that ${\rm SNR}>1$ you can find the observable parameter space of your GW signal. In the case of GWs induced by peaked primordial spectrum you can constraint the amplitude of the spectrum vs the peak wavenumber. And, with this, we end the lectures.

\begin{tcolorbox}[title=\bf{Try it!},coltitle=black,colback=white,colframe=orange!50]
Download the latest version of \href{https://github.com/Lukas-T-W/SIGWfast/releases}{SIGWfast} by Lukas T. Witkowski and compute the resulting spectrum of GWs induced by your favorite shape for the primordial spectrum of scalar fluctuations, and your favorite equation of state of the primordial universe after inflation. Check also the companion paper in the \href{https://arxiv.org/abs/2209.05296}{arXiv}.
\tcblower
\textbf{Warning:} if you aim to compute the induced GWs for very sharp spectra, you will have to increase the resolution (controlled by the parameters ns1 and ns2) in the file called \textsc{sdintegral.py} in the libraries folder.
\end{tcolorbox}

\begin{tcolorbox}[title=\bf{Try it!},coltitle=black,colback=white,colframe=orange!50]
You can plot the GW sensitivity curves and forecast constraints on the amplitude of the primordial spectrum by using the method explained by Kai Schmitz \href{https://arxiv.org/abs/2002.04615}{here}. Kai also has a repository with the sensitivity curves at \href{https://zenodo.org/record/3689582}{zenodo}.
\end{tcolorbox}

\section*{Acknowledgments} 
I am supported by the DFG under the Emmy-Noether program grant no. DO 2574/1-1, project number 496592360. I would also like to thank the organizers of the PhD school in Barcelona for inviting me to give these lectures as well as the participants for the interesting discussions and feedback during the lectures.


\chapter*{Appendices}
\addcontentsline{toc}{chapter}{Appendices}
\markboth{Appendices}{Appendices}

\appendix

\renewcommand{\thesection}{\Alph{section}}

\section{Additional formulas \label{app:formulasuseful}}

Here I will briefly show where the effective degrees of freedom appear in the equations. More detailed explanations can be found, e.g., in Baumann's lecture notes \url{http://cosmology.amsterdam/education/cosmology/}.

The main point is that, considering the standard model of particles, as we raise the temperature of the universe, more and more particles become relativistic and contribute to the radiation component. You may remember from Statistical Physics class that one can use the canonical ensemble to describe a collection of particles at a temperature bath $T$. Well, you can do the same in cosmology. For relativistic particles, you can relate the temperature of the fluid (or gas) to the number, energy, entropy density, pressure, etcetera. Fermions and bosons have a different coefficient in this relation. But, in the end, we can count how many of each we have and how much they contribute. In particular the energy and entropy density are given by
\begin{align}
\rho=\frac{\pi^2}{30}g_{\rho}(T)T^4\,
\quad{\rm and}\quad
s=\frac{2\pi^2}{45}g_{s}(T)T^3\,.
\end{align}
At very high temperatures, i.e., $T\gg 100\,{\rm GeV}$, when all standard model particles are relativistic one has $g_{\rho}\approx g_{s}\approx 106.75$. At the time of matter-radiation equality and at present we took $g_\rho(T_{\rm eq})\approx g_{\rho}(T_{0})\approx 3.38$ and $g_{s}(T_{\rm eq})\approx g_{s}(T_{0})\approx 3.94$. See Ref.~\cite{Husdal:2016haj} for a review on $g_{\rho}$ and $g_{s}$ at different times.

If radiation dominates the universe we can use the Friedmann equation ($3H^2M_{\rm pl}^2=\rho$) to write
\begin{align}\label{eq:HofT}
H=\frac{\pi}{3\sqrt{10}M_{\rm pl}}g_\rho^{1/2}(T)T^2\,.
\end{align}
Most important is that entropy is conserved! In an expanding universe this means that $s\propto a^{-3}$. We can then use entropy conservation to relate the temperature at two different times with the scale factor of the universe at those time. Doing so we arrive at
\begin{align}
\frac{a}{a_\star}=\frac{T_\star}{T}\left(\frac{g_{s}(T_\star)}{g_{s}(T)}\right)^{1/3}\,.
\end{align}

\section{Scalar vector tensor decomposition \label{app:decomposition}}

Take a 3 dimensional rank 2 symmetric tensor (this means it has 6 independent components), call it $H_{ij}$. Let's identify each independent component. You can first consider the trace, say $H=\delta^{ij}H_{ij}$. I assumed a flat FLRW so the background spatial metric is $\delta_{ij}$. Then, you can for example, take two derivatives of $H_{ij}$ minus the trace, e.g.
\begin{align}
S=\frac{\partial^i\partial^j}{\Delta}(H_{ij}-\frac{1}{3}\delta_{ij}H)\,,
\end{align}
where $\Delta=\partial^k\partial_k$. Note that $1/\Delta$ is formally non-local and it makes sense when doing a Fourier expansion. $S$ and $H$ have no indices and are scalars. We can then subtract the scalar part and identify the non fully transverse components (i.e. that do not vanish after taking a derivative). This can be done with the transverse projector given by
\begin{align}
P_{ij}=\delta_{ij}-\frac{\partial_i\partial_j}{\Delta}\,.
\end{align}
It is easy to check that $\partial^iP_{ij}=0$. We then find that the resulting component reads
\begin{align}
F^k=P^{ik}\left(H_{ij}-\frac{1}{3}\delta_{ij}H-(\partial_{i}\partial_{j}-\frac{1}{3}\delta_{ij}\Delta)S\right)\,,
\end{align}
and as it has one index is a vector. It has two independent components because it is transverse, i.e. $\partial_kF^k=0$. The remaining independent components are the transverse-traceless piece. We can select it with the transverse-traceless projector, namely
\begin{align}
\widehat{TT}_{ijkl}=P_{ik}P_{jl}-\frac{1}{2}P_{ij}P_{kl}\,.
\end{align}
Then, the transverse-traceless component is
\begin{align}
h_{kl}=\widehat{TT}^{ij}_{kl}\left(H_{ij}-\frac{1}{3}\delta_{ij}H-(\partial_{i}\partial_{j}-\frac{1}{3}\delta_{ij}\Delta)S-\partial_{(i}F_{j)}\right)\,.
\end{align}
These are the so-called tensor modes and are the ones connected to GWs. They satisfy $\delta^{ij}h_{ij}=\partial^ih_{ij}=0$ and have two independent components. So for $H_{ij}$ we have 2 scalars, 1 vector and 1 tensor. Note that this decomposition works very well for linear perturbations but it mixes if we go to higher order in perturbation theory.\\

\begin{tcolorbox}[title=\bf{Exercise},colback=white,colframe=black!50]
Check that indeed $\widehat{TT}_{ijkl}$ satisfies the transverse-traceless conditions.
\end{tcolorbox}

\section{Polarization tensors  \label{app:polarization}}
Here I provide explicit expression for the polarization tensors. I choose to work in spherical coordinates and assume that the GW propagates along
\begin{align}
\mathbf{k}=k(\sin\theta_k\cos\chi_k,\sin\theta_k\sin\chi_k,\cos\theta_k)\,,
\end{align}
where $\theta_k$ and $\chi_k$ respectively are the polar and azimuthal angles. With this choice, a pair of orthonormal polarization vectors are
\begin{align}
\mathbf{e}(\mathbf{k})&=(\cos\theta_k\cos\chi_k,\cos\theta_k\sin\chi_k,-\sin\theta_k)\,,\\
\bar{\mathbf{e}}(\mathbf{k})&=(-\sin\chi_k,\cos\chi_k,0)\,.
\end{align}
Then we build the polarization tensors for plus $+$ and cross $\times$ polarization as
\begin{align}
e^{+}_{ij}(\mathbf{k})&=\frac{1}{\sqrt{2}}\left[e_{i}(\mathbf{k})e_{j}(\mathbf{k})-\bar e_{i}(\mathbf{k})\bar e_{j}(\mathbf{k})\right]\,,\\
e^{\times}_{ij}(\mathbf{k})&=\frac{1}{\sqrt{2}}\left[e_{i}(\mathbf{k})\bar e_{j}(\mathbf{k})+\bar e_{i}(\mathbf{k}) e_{j}(\mathbf{k})\right]\,.
\end{align}
which satisfy
\begin{align}
e^{+}_{ij}(\mathbf{k})e^{+ ij}(-\mathbf{k})&=1\quad,\quad e^{\times}_{ij}(\mathbf{k})e^{\times ij}(-\mathbf{k})=1\quad,\quad e^{+}_{ij}(\mathbf{k})e^{\times ij}(-\mathbf{k})=0\,,\nonumber\\
\delta_{ij}e^{+ij}(\mathbf{k})&=\delta_{ij}e^{\times ij}(\mathbf{k})=k_ie^{+ij}(\mathbf{k})=k_ie^{\times ij}(\mathbf{k})=0\,.
\end{align}
For the right and left polarizations we then have
\begin{align}
e^{R}_{ij}(\mathbf{k})=\tfrac{1}{\sqrt{2}}(e^{+}_{ij}(\mathbf{k})+ie^{\times}_{ij}(\mathbf{k}))\quad{\rm and}\quad e^{L}_{ij}(\mathbf{k})=\tfrac{1}{\sqrt{2}}(e^{+}_{ij}(\mathbf{k})-ie^{\times}_{ij}(\mathbf{k}))\,.
\end{align}

When contracted with a momentum $\mathbf{q}$ given by
\begin{align}
\mathbf{q}=k(\sin\theta_q\cos\chi_q,\sin\theta_q\sin\chi_q,\cos\theta_q)\,,
\end{align}
we find that
\begin{align}
e^{R}_{ij}(\mathbf{k})q_iq_j&=\frac{1}{2}q^2e^{-2i\chi_q}\sin^2\theta_q\,,\\
e^{L}_{ij}(\mathbf{k})q_iq_j&=\frac{1}{2}q^2e^{2i\chi_q}\sin^2\theta_q\,.
\end{align}

\section{Gauge transformations \label{app:gaugetransformations}}

Here I provide some useful way to compute the gauge transformations, i.e. how perturbations variables change under a change of coordinates. It is convenient to use the non-linear gauge transformations under the exponential mapping \cite{Malik:2008im} (which are actually related to Hamiltonian flows). These are given by
\begin{align}
\hat {\tilde g}_{\mu\nu}={\rm e}^{{\cal L}_\xi}{\tilde g}_{\mu\nu}={\tilde g}_{\mu\nu}+{\cal L}_\xi{\tilde g}_{\mu\nu}+\frac{1}{2}{\cal L}^2_\xi{\tilde g}_{\mu\nu}+...\,,
\end{align}
where $\hat {\tilde g}_{\mu\nu}$ is the metric in one gauge and ${\tilde g}_{\mu\nu}$ in another. ${\cal L}_\xi$ is the lie derivative along a direction $\xi^\mu$ which can roughly be thought of the coordinate transformation $\hat {\tilde x}^\mu=\tilde x^\mu+\xi^\mu$. A nice property of the exponential mapping is that it is easy to expand formally the conformal transformation, namely
\begin{align}
{\rm e}^{{\cal L}_\xi}{\tilde g}_{\mu\nu}={\rm e}^{{\cal L}_\xi}\left[a^2{ g}_{\mu\nu}\right]&=\left({\rm e}^{{\cal L}_\xi}a^2\right)\left({\rm e}^{{\cal L}_\xi}{g}_{\mu\nu}\right)\nonumber\\&=\left(a^2+{\cal L}_\xi a^2+\frac{1}{2}{\cal L}^2_\xi a^2+...\right)\left(g_{\mu\nu}+{\cal L}_\xi g_{\mu\nu}+\frac{1}{2}{\cal L}^2_\xi g_{\mu\nu}+...\right)\,.
\end{align}
In this way, when we compute the gauge transformation of the Einstein tensor, the  conformal factor that will come from the Lie derivative of the scale factor can be factor out using a conformal transformation. This will be particularly useful to carry the results of Minkowski space-time to a spatially flat FLRW cosmology.

I also provide the expressions for the Lie derivatives of scalars, vectors and tensors.  The first Lie derivatives are given by
\begin{align}
&{\cal L}_\xi A=\xi^\alpha\partial_\alpha A\,,\\
&{\cal L}_\xi B_\mu=\xi^\alpha\partial_\alpha B_\mu+B_\alpha\partial_\mu \xi^\alpha\,, \\
&{\cal L}_\xi C_{\mu\nu}=\xi^\alpha\partial_\alpha C_{\mu\nu}+2C_{\alpha(\mu}\partial_{\nu)}\xi^{\alpha}\,,
\end{align}
where $A$ is an arbitrary scalar, $B$ is an arbitrary vector and $C$ is an arbitrary tensor. The second Lie derivative reads
\begin{align}
&{\cal L}^2_\xi A=\xi^\beta\partial_\beta\left(\xi^\alpha\partial_\alpha A\right)\,,\\
&{\cal L}^2_\xi B_\mu=\xi^\beta\partial_\beta\left(\xi^\alpha\partial_\alpha B_\mu+B_\alpha\partial_\mu \xi^\alpha\right)+\xi^\alpha\partial_\alpha u_\beta\partial_\mu\xi^\beta+B_\alpha\partial_\beta \xi^\alpha\partial_\mu \xi^\beta \,,\\
&{\cal L}^2_\xi C_{\mu\nu}=\xi^\beta\partial_\beta\left(\xi^\alpha\partial_\alpha C_{\mu\nu}+2C_{\alpha(\mu}\partial_{\nu)}\xi^{\alpha}\right)+2\xi^\alpha \partial_\alpha C_{\beta(\mu}\partial_{\nu)}\xi^\beta\nonumber\\&\hspace*{2cm}+2C_{\alpha\beta}\partial_{(\mu}\xi^\alpha\partial_{\nu)}\xi^\beta+2\partial_\beta\xi^\alpha C_{\alpha(\mu}\partial_{\nu)}\xi^\beta\,.
\end{align}

Now we can check how the metric transforms. First we have that
\begin{align}
&{\cal L}_\xi a^2=2a^2\xi^0{\cal H}\,,\\
&{\cal L}^2_\xi a^2=2a^2\left\{(\xi^0)^2\left[2{\cal H}^2+{\cal H}'\right]+{\cal H}\xi^\alpha\partial_\alpha \xi^0\right\}\,.
\end{align}
And, for the metric we have
\begin{align}
&{\cal L}_\xi{\eta}_{\mu\nu}=2\partial_{(\mu}\xi_{\nu)}\,,\\
&{\cal L}_\xi{h}_{\mu\nu}=\xi^\alpha\partial_\alpha h_{\mu\nu}+2h_{\alpha(\mu}\partial_{\nu)}\xi^{\alpha}\,,\\
&{\cal L}^2_\xi{\eta}_{\mu\nu}=2\xi^\alpha\partial_\alpha\partial_{(\mu}\xi_{\nu)}+2\partial_{(\mu}\xi^\alpha\partial_{\nu)}\xi_\alpha+2\partial_\alpha\xi_{(\mu}\partial_{\nu)}\xi^\alpha\,.
\end{align}

Using all the above we find that
\begin{align}
h_{\mu\nu}\to h_{\mu\nu}+\delta^{(1)} h_{\mu\nu}+\delta^{(2)} h_{\mu\nu}\,,
\end{align}
where
\begin{align}
&\delta^{(1)} h_{\mu\nu}=2\partial_{(\mu}\xi_{\nu)}+2\eta_{\mu\nu}\xi^0{\cal H}\,,\\
&\delta^{(2)} h_{\mu\nu}=\xi^\alpha\partial_\alpha h_{\mu\nu}+
2h_{\alpha(\mu}\partial_{\nu)}\xi^{\alpha}
+\xi^\alpha\partial_\alpha\partial_{(\mu}\xi_{\nu)}+\partial_{(\mu}\xi^\alpha\partial_{\nu)}\xi_\alpha+\partial_\alpha\xi_{(\mu}\partial_{\nu)}\xi^\alpha\nonumber\\&
\hspace{1cm}+2\left[h_{\mu\nu}+2\partial_{(\mu}\xi_{\nu)}\right]\xi^0{\cal H}+\eta_{\mu\nu}\left[(\xi^0)^2\left[2{\cal H}^2+{\cal H}'\right]+{\cal H}\xi^\alpha\partial_\alpha \xi^0\right]\,.
\end{align}

\chapter*{References}
\addcontentsline{toc}{chapter}{References}

\pagestyle{plain}

\printbibliography[heading=none]

\end{document}